\documentclass[aps,prb,reprint,noeprint,superscriptaddress,nobibnotes,longbibliography]{revtex4-2}
\pdfoutput=1
\usepackage[utf8]{inputenc}
\usepackage[english]{babel}
\usepackage{microtype}
\usepackage{amsmath,amssymb,amsfonts}
\usepackage{mathtools}
\usepackage{braket}
\usepackage{bm}
\usepackage{graphicx}
\usepackage{booktabs}

\usepackage[colorlinks=true,unicode=true,pdfborder={0 0 0},allcolors=blue
]{hyperref}

\renewcommand{\vec}[1]{\bm{#1}}
\newcommand{\im}{\mathrm{i}}
\newcommand{\e}{\mathrm{e}}

\renewcommand{\Re}{\operatorname{\mathsf{Re}}}
\newcommand{\vk}{{\vec k}}
\newcommand{\vq}{{\vec q}}
\newcommand{\vQ}{{\vec{Q}}}
\newcommand{\VV}[2]{\begin{pmatrix}#1\\#2\end{pmatrix}}
\newcommand{\VVT}[2]{\begin{pmatrix}#1, & #2\end{pmatrix}}
\newcommand{\VVV}[3]{\begin{pmatrix}#1\\#2\\#3\end{pmatrix}}
\newcommand{\VVVT}[3]{\begin{pmatrix}#1, & #2, & #3\end{pmatrix}}
\newcommand{\MM}[4]{\begin{pmatrix}#1 & #2 \\ #3 & #4\end{pmatrix}}
\newcommand{\tr}{\operatorname{tr}}

\begin{document}

\title{Phase signatures
in third-harmonic response\\of Higgs and coexisting modes superconductors}

\author{Lukas Schwarz}
\affiliation{Max Planck Institute for Solid State Research,
70569 Stuttgart, Germany}

\author{Rafael Haenel}
\affiliation{Max Planck Institute for Solid State Research,
70569 Stuttgart, Germany}
\affiliation{Stewart Blusson Quantum Matter Institute, University of British Columbia, Vancouver V6T 1Z4, Canada}

\author{Dirk Manske}
\affiliation{Max Planck Institute for Solid State Research,
70569 Stuttgart, Germany}

\date{\today}

\begin{abstract}
Third-harmonic generation (THG) experiments on superconductors
can be used to investigate collective excitations like the amplitude mode
of the order parameter known as Higgs mode.
These modes are visible due to resonances in the THG signal
if the driving frequency matches the energy of the mode.
In real materials multiple modes can exist
giving rise to additional THG contributions,
such that it is difficult to unambiguously interpret the results.
In this paper, we additionally analyze the phase of the THG signal,
which contains microscopic details beyond classical resonances
as well as signatures of couplings between modes
which are difficult to observe in the amplitude alone.
We investigate how the Higgs mode,
impurities or Coulomb interaction affects the phase response
and consider exemplary two systems with additional modes.
We argue that extracting this phase information could be valuable in future experiments.
\end{abstract}

\maketitle

\section{Introduction}
Recent progress in ultrafast THz laser technology lead to an increasing interest
in studying collective excitations of superconducting systems.
Especially the investigation of the amplitude (Higgs) mode of superconductors
lead to a new emerging field termed Higgs spectroscopy,
where the Higgs mode is used as a probe for intrinsic properties of the system
\cite{Matsunaga2013,Matsunaga2014,Chu2020,Schwarz2020a}.
Intrinsically, a superconductor possesses two collective modes
due to the spontaneous $U(1)$ symmetry breaking:
An amplitude oscillation of the order parameters, known as the Higgs mode
and a phase oscillation, known as the Goldstone mode \cite{Varma2002,Pekker2015}.
The Goldstone mode is a massless mode
in the long-wave limit for uncharged systems.
However, the Anderson-Higgs mechanism in a charged superconductor
shifts its energy to the plasma energy.
The Higgs mode is a massive mode
with an energy at the quasiparticle energy $2\Delta$
and thus is a low-energy excitation in the range of THz frequency.

Experiments to excite the Higgs mode are usually performed in either of two ways.
One option is to quench the system
with an ultrafast, single-cycle THz pump pulse
to abruptly change the system's parameter and bring it out of equilibrium
\cite{Papenkort2007,Krull2014}.
The order parameter starts to automatically oscillate
around its new equilibrium state with the Higgs mode frequency.
This oscillation is experimentally measured in a pump-probe geometry,
where the probe pulse scans the dynamics of the system
with a variable time-delay after the pump pulse \cite{Matsunaga2013}.

The second option is to drive the system periodically
with a multi-cycle THz pulse at frequency $\Omega$.
This enforces the order parameter
to oscillate with twice the driving frequency $2\Omega$
due to the quadratic excitation process \cite{Tsuji2015,Cea2016,Schwarz2020}.
Furthermore, this leads to a third-harmonic generation (THG) process,
which can be measured in the transmitted electric field.
Tuning the driving frequency into resonance
with the Higgs mode energy, i.e. $2\Omega = 2\Delta$,
a resonance peak is visible in the signal.
This can be achieved either by varying the driving frequency
or, as it is currently done experimentally,
by changing the value of the order parameter $\Delta(T)$
by sweeping the temperature $T$.
The resonance can be used as a signature for the collective Higgs mode
as it was demonstrated for the $s$-wave superconductor NbN \cite{Matsunaga2013,Matsunaga2017}.

In many materials,
more complicated effects may arise resulting from coexisting modes
additionally contributing to the THG signal.
Examples include quasiparticle excitations \cite{Cea2016},
Leggett modes in multiband systems \cite{Murotani2017,Murotani2019,Haenel2021},
Josephson-Plasma modes in layered systems \cite{Gabriele2021},
Bardasis-Schrieffer modes in systems with subleading pairing channels \cite{Mueller2019},
coexisting CDW fluctuations \cite{Cea2014}
or generally phonon and magnon excitations.

Our work is motivated by a recent THG experiment on several cuprates \cite{Chu2020}.
The experiment revealed an interesting phase signature
containing antiresonance behavior
which cannot be explained by the excitation of a single collective mode.
In our work, we therefore take into account the existence of another mode
and we investigate the THG signal for such systems,
where we concentrate on the phase of the $3\Omega$ oscillations,
which was not discussed theoretically so far.
As it is well known, a driven oscillator shows an abrupt phase change
if the driving frequency is varied across the eigenfrequency of the system.
Furthermore, multiple coupled oscillators show an antiresonance,
resulting from the interplay of driving force and coupling,
where a minimum in the driving amplitude occurs
accompanied by a negative phase change.
This behavior occurs in many physical systems
and we show that it is also visible in the THG response
following from a microscopic calculation with coupled modes.
Yet, the response is more complex compared to a classical model
as modified mode propagators and susceptibility terms occur.

After a general analysis, we
provide two detailed microscopic calculations
of a two mode scenario.
First, a coupling of the Higgs mode to a charge density wave
and second a Higgs mode with a coexisting Bardasis-Schrieffer mode.
With this paper, we propose that analyzing the phase of the THG signal
in addition to the amplitude
yields additional information
valuable for understanding the interplay of superconductivity and other modes.

The paper is organized as follows. In Sec.~\ref{sec:single-mode},
we generally discuss the phase response of a single driven oscillator.
We start from a classical model,
proceed to a phenomenological Ginzburg-Landau theory for superconductivity and
finally show a microscopic calculation in an effective action formalism.
In Sec.~\ref{sec:multiple-modes},
we extend the single mode analysis to a second mode.
We start again with a classical model
and then discuss the general features of a microscopic theory.
In Sec.~\ref{sec:higgs-cdw},
we explicitly calculate the THG response of a superconducting system
with coexisting CDW.
In Sec.~\ref{sec:higgs-bs},
we explicitly calculate the THG response of a superconducting system
with Bardasis-Schrieffer modes.
Finally, we summarize and conclude in Sec.~\ref{sec:conclusion}

\section{Phase signature of a single mode}
\label{sec:single-mode}
Before studying the full microscopic
quantum mechanical model for superconductors and its collective modes,
let us first consider a simple classical system.
This will allow us to define and observe the crucial features
which are important for the later discussion.
Hereby, we investigate classical driven oscillators
which represent the collective modes of the system.

\subsection{Harmonic oscillator}
\label{sec:single-ho}
It is well known that a driven harmonic oscillator has a characteristic
amplitude and phase response which depends on the driving frequency.
With the eigenfrequency $\omega_0$, damping factor $\gamma$, driving amplitude
$F_0$ and driving frequency $\Omega$, the equation of motion
for the displacement $x(t)$ reads
\begin{align}
    \ddot x(t) + \omega_0^2 x(t) + \gamma \dot x(t) = F_0 \cos(\Omega t)\,.
\end{align}
The steady-state solution can be written as
$x(t) = A \cos(\Omega t - \phi)$,
where the frequency-dependent amplitude $A$ and phase $\phi$ are given by
\begin{subequations}
\begin{align}
    A(\Omega) &= \frac{F_0}
        {\sqrt{(\omega_0^2 - \Omega^2)^2 + \gamma^2\Omega^2}}\,,\\
    \phi(\Omega) &= \tan^{-1}\left(
        \frac{\gamma\Omega}{\omega_0^2-\Omega^2}
    \right)\,.
\end{align}
\label{eq:ho_solution}%
\end{subequations}
One observes that the amplitude has a resonance peak at $\Omega = \omega_0$
which is accompanied by an abrupt phase change from $0$ to $\pi$.
Thus, the oscillation is in-phase
with the driving frequency below the resonance and
lags behind with opposite phase above the resonance.

\begin{figure}[t]
    \centering
    \includegraphics[width=\columnwidth]{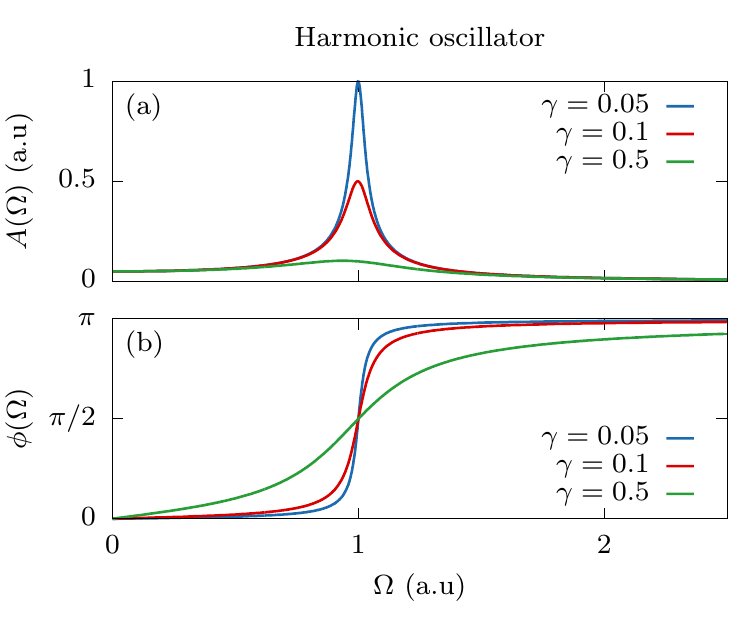}
    \caption{\label{fig:single_mode_ho}%
    (a) Amplitude and (b) phase of a driven harmonic oscillator
    for different damping $\gamma$ according to Eq.~\eqref{eq:ho_solution}
    with $F_0 = 1$ and $\omega_0 = 1$.}
\end{figure}%

The amplitude and phase is plotted in Fig.~\ref{fig:single_mode_ho}
for different damping values $\gamma$.
While for small damping a pronounced resonance peak is visible in the amplitude,
for large damping, the resonance peak is heavily suppressed and broadened.
In contrast, the phase still shows a phase change from $0$ to $\pi$,
even though it is broadened as well.
This means that both amplitude and phase have a signature of the resonance,
yet the phase change signature
is more robust against the influence of damping.
Hence, in a strongly damped system with suppressed resonance peak,
the eigenmode would still be identifiable via the phase signature.

\subsection{Ginzburg-Landau model}
\label{sec:ginzburg-landau}
Let us investigate now
whether we can observe such a behavior
for THz-driven collective modes in superconductors as well.
Hereby, the oscillator corresponds to a collective mode
which is driven by a THz light field.
In the experiment, the driven collective mode is not measured directly.
Instead the induced current proportional to the transmitted electric field is recorded.

As a first step, we investigate the phenomenological Ginzburg-Landau model,
where we will consider amplitude and phase fluctuations.
The time-dependent Lagrangian of a superconductor coupled to a gauge field
is given by
\begin{align}
    \mathcal L = (D_\mu \psi)^* (D^\mu \psi)
    - V(\psi) - \frac 1 4 F_{\mu\nu}F^{\mu\nu}\,,
\end{align}
where $\psi$ is the superconducting order parameter,
$D_\mu = \partial_\mu + \im e A_\mu$ the covariant derivative
with the four-vectors $\partial^\mu = (\partial_t,-\nabla)$
and $A_\mu = (\Phi,-\vec A)$
and electromagnetic field tensor
$F_{\mu\nu} = \partial_\mu A_\nu - \partial_\nu A_\mu$
in units where $c=1$.
In principle, the Lagrangian could also contain additional linear derivative terms.
Yet, we assume perfect particle-hole symmetry,
such that the time dynamics of a superconductor is
described only by a second-order derivative term
\cite{Varma2002,Pekker2015}.
The potential $V(\psi) = \alpha|\psi|^2 + \frac{\beta}{2}|\psi|^4$
is the free energy of a superconductor with $\beta > 0$
and $\alpha = \alpha_0(T-T_c)$
such that for $T < T_c$ the potential takes the form of a Mexican hat
with the ground state $\psi_0 = \sqrt{-\alpha/\beta}$.
We introduce amplitude (Higgs) fluctuations $H(\vec r,t)$
and phase (Goldstone) fluctuations $\theta(\vec r,t)$
via
\begin{align}
    \psi(\vec r, t) &= (\psi_0 + H(\vec r,t))\e^{\im\theta(\vec r,t)}\,,
\end{align}
and choose a gauge $A_\mu \rightarrow A_\mu + \frac 1 e \partial_\mu \theta$
and $\psi \rightarrow \psi \e^{-\im\theta}$.
Then, the Lagrangian up to second order in the fluctuations reads
\begin{align}
    \mathcal L &= (\partial_\mu H)(\partial^\mu H) + 2\alpha H^2
        - \frac{1}{4} F_{\mu\nu}F^{\mu\nu}
    \notag\\&\qquad
        + e^2\psi^2_0 A_\mu A^\mu + 2e^2 \psi_0 A_\mu A^\mu H\,.
\end{align}
Hereby, the phase fluctuations are removed from the Lagrangian
by the chosen gauge and are implicitly included
in the longitudinal component of the transformed gauge field $A_\mu$
which obtains an additional mass term $\propto A_\mu A^\mu$.
This effect is known as the Anderson-Higgs mechanism \cite{Anderson1963}.
Calculating the equations of motion for the Higgs mode $H$,
neglecting spatial fluctuations for $\vq\rightarrow 0$
and choosing a gauge with $\Phi = 0$,
yields
\begin{align}
    \partial_t^2 H(t) - 2\alpha H(t) = -e^2\psi_0 A(t)^2\,.
\end{align}
The dynamics of the Higgs oscillations is governed by
a harmonic oscillator with frequency $\omega_0 = \sqrt{-2\alpha}$.
The driving term is quadratic in the vector potential $A(t)$.
With a periodic light field $A(t) = A_0\cos(\Omega t)$,
the system is effectively driven by $2\Omega$
such that the resonance in the system occurs at $2\Omega = \omega_0$.
Thus, on a phenomenological level,
the collective Higgs oscillations of a superconductor
and its amplitude and phase signature
is exactly described by the classical model discussed before.
The measured transmitted field is described by the induced current given by \cite{Tsuji2015}
\begin{align}
    j(t) &= \frac{\partial \mathcal L}{\partial A}
    = -2e^2\psi_0^2 A(t) - 4e^2\psi_0 A(t) H(t)
    \label{eq:j_tdgl}\,.
\end{align}
A nonlinear third-harmonic component in the current is induced
as $A(t)\cdot H(t) \propto \cos(3\Omega t - \phi) + \ldots$
The resonance behavior of the amplitude and phase in the current $j(t)$
is directly given by the Higgs response $H(t)$.

\subsection{Microscopic BCS model}
\label{sec:bcs}
While in the phenomenological model the coupling of light to the system
contains no further details,
in a microscopic model additional effects
with frequency-dependent susceptibilities occur.
Furthermore, there are quasiparticles in the microscopic model
which render the Higgs mode less stable due to the additional decay channel.

\begin{figure}[t]
    \centering
    \includegraphics[width=\columnwidth]{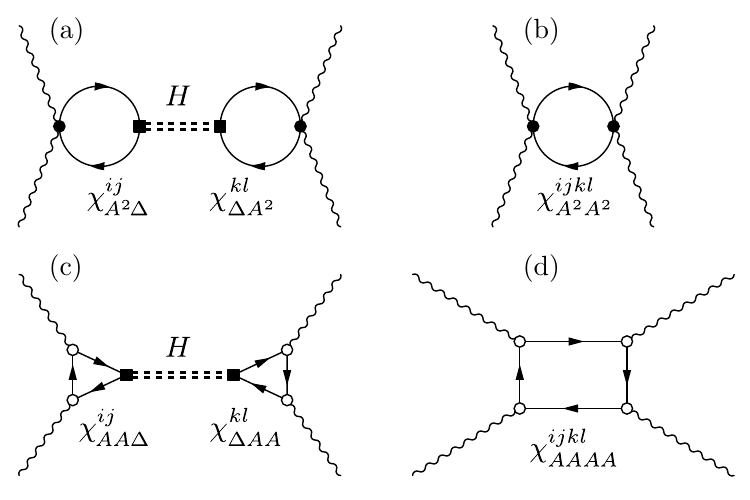}
    \caption{\label{fig:single_mode_diagram}%
    Diagrammatic representation of the effective action
    in the (a),(b) clean limit according to Eq.~\eqref{eq:k4_kernel}
    and in the (c),(d) dirty limit according to Eq.~\eqref{eq:chi_dirty_limit}.
    (a) Diamagnetic Higgs excitation.
    (b) Diamagnetic quasiparticle excitation.
    (c) Paramagnetic Higgs excitation.
    (d) Paramagnetic quasiparticle excitation.
    The wiggly lines represent the vector potential $A$,
    the solid lines the BCS Green's function $G_0$
    and the double dashed line the Higgs propagator $H$.
    The filled square vertex corresponds to $f_\vk \tau_1$,
    the filled circle vertex to $\partial_{ij}^2\epsilon_\vk \tau_3$
    and the empty circle vertex to $\partial_i\epsilon_\vk \tau_0$.}
\end{figure}

To address these effects, we proceed to the full microscopic theory
using an effective action approach \cite{Altland2010,Cea2016}.
The BCS Hamiltonian reads
\begin{align}
    H_{\mathrm{BCS}}(t) &= \sum_{\vk,\sigma} \epsilon_\vk c_{\vk,\sigma}^\dagger c_{\vk,\sigma}
        - \sum_{\vk,\vk'} V_{\vk,\vk'} c_{\vk,\uparrow}^\dagger
            c_{-\vk,\downarrow}^\dagger c_{-\vk',\downarrow} c_{\vk',\uparrow}
        \notag\\&\qquad
        + \frac 1 2 \sum_{\vk,\sigma} \sum_{i,j} \partial_{ij}^2\epsilon_\vk
        A_i(t) A_j(t) c_{\vk,\sigma}^\dagger c_{\vk,\sigma}\,.
    \label{eq:bcs_hamiltonian}
\end{align}
Hereby, $\epsilon_\vk = \xi_\vk - \epsilon_{\mathrm F}$
is the electron dispersion $\xi_\vk$
measured relative to the Fermi level $\epsilon_{\mathrm F}$
and $c_{\vk,\sigma}^\dagger$ or $c_{\vk,\sigma}$
the electron creation or annihilation operators.
The separable BCS pairing interaction
is given by $V_{\vk,\vk'} = V f_\vk f_{\vk'}$ with pairing strength $V$
and symmetry $f_\vk$.
A coupling to light represented by the vector potential $\vec A(t)$
is realized by minimal coupling
$\epsilon_\vk \rightarrow \epsilon_{\vk-\vec A(t)}$.
An expansion in powers of $\vec A(t)$
yields the lowest non-vanishing diamagnetic coupling term shown above,
while the linear paramagnetic coupling $\propto \partial_i A_i(t)$
vanishes due to parity symmetry.
In the expression,
we have introduced the short-hand notation $\partial_{ij}^2 = \partial_{k_ik_j}^2$.
Here,
we initially neglect long-ranged Coulomb interaction
and the coupling to phase fluctuations which is important in real materials.
We will show later in Sec.~\ref{sec:coulomb}
that including Coulomb interaction does not affect the phase signature.
The action of the system in imaginary time $\tau$ is given by
\begin{align}
    S &= \int_0^\beta \mathrm d\tau \left(
        \sum_{\vk,\sigma} c_{\vk,\sigma}^\dagger(\tau)
        \partial_\tau c_{\vk,\sigma}(\tau) + H(\tau)
        \right)\,.
\end{align}
We perform a Hubbard-Stratonovich transformation
introducing the bosonic field $\Delta$,
with amplitude fluctuations $\Delta(t) = \Delta + \delta\Delta(t)$.
After integration of fermions,
we split the action in a mean-field and fluctuating part,
which we expand up to fourth order in A.
For more details about the calculation
see Appendix~\ref{app:higgs}.
The effective action with Matsubara frequencies $\im\omega_m$
in fourth-order of the vector potential reads
\begin{align}
    S^{(4)} &= \frac 1 2 \frac 1 \beta \sum_{\im\omega_m}
        \delta\Delta(-\im\omega_m) H^{-1}(\im\omega_m) \delta\Delta(\im\omega_m)
        \notag\\&\qquad
        - 2\delta\Delta(-\im\omega_m) \sum_{i,j}
        \chi_{\Delta A^2}^{ij} A_{ij}^2(\im\omega_m)
        \notag\\&\qquad
        + \sum_{ijkl}
        A_{ij}^2(-\im\omega_m) \chi_{A^2A^2}^{ijkl}(\im\omega_m) A_{kl}^2(\im\omega_m)\,.
\end{align}
Hereby, $H^{-1}(\im\omega_m)$ is the inverse Higgs propagator defined as
the renormalized pairing interaction $V$
\begin{align}
    H^{-1}(\im\omega_m) &= \frac 2 V + \chi_{\Delta\Delta}(\im\omega_m)\,.
\end{align}
The susceptibilities are given by
\begin{subequations}
\begin{align}
    \chi_{\Delta\Delta}(\im\omega_m) &= \sum_\vk f_\vk^2 X_{11}(\vk,\im\omega_m)\,,\\
    \chi_{\Delta A^2}^{ij}(\im\omega_m) &=
        \sum_\vk f_\vk \, \frac 1 2 \partial_{ij}^2 \epsilon_\vk \,
        X_{13}(\vk,\im\omega_m)\,,\\
    \chi_{A^2 A^2}^{ijkl}(\im\omega_m) &= \sum_\vk \frac 1 4 \partial_{ij}^2 \epsilon_\vk \,
        \partial_{kl}^2 \epsilon_\vk \,
        X_{33}(\vk,\im\omega_m)
\end{align}
\end{subequations}
with
\begin{align}
    X_{\alpha\beta}(\vk,\im\omega_m) &= \frac 1 \beta \sum_{\im\omega_n}
        \tr[
        G_0(\vk,\im\omega_n) \tau_\alpha G_0(\vk,\im\omega_m+\im\omega_n) \tau_\beta
        ]
    \label{eq:XX}
\end{align}
and the BCS Green's function
$G_0^{-1}=\im\omega_m \tau_0 - \epsilon_\vk \tau_3 + \Delta_\vk \tau_1$
where $\tau_i$ are Pauli matrices.
The indices $\Delta$ and $A^2$ in the susceptibilities
represent the vertices,
i.e. the coupling to the Higgs propagator via $f_\vk \tau_1$
or the coupling to light via $\partial_{ij}^2\epsilon_\vk \tau_3$, respectively.
Integrating out the amplitude fluctuations
and after analytic continuation $\im\omega_m\rightarrow \omega + \im0^+$
one obtains
\begin{align}
    S^{(4)} &= \frac 1 2 \int \mathrm d\omega
        \sum_{ijkl} \Big(
        \chi_{\Delta A^2}^{ij}(-\omega)
            \chi_{\Delta A^2}^{kl}(\omega) H(\omega)
        \notag\\&\quad
        + \chi_{A^2A^2}^{ijkl}(\omega)
        \Big)
        A_{ij}^2(-\omega)A_{kl}^2(\omega)\,.
    \label{eq:action-clean}
\end{align}
There are two contributions in the action,
one containing the Higgs oscillations and one the quasiparticle response.
These contributions are shown diagrammatically
in Fig.~\ref{fig:single_mode_diagram}(a) and (b).

For simplicity, we will only consider linear-polarized light in $x$-direction,
such that we can neglect the polarization indices in the following.
With this, the third-harmonic response is given by
\begin{align}
    j^{(3)}(3\Omega) &= -\left.\frac{\delta S^{(4)}}{\delta A(-\omega)}\right|_{3\Omega}
        \propto \chi_H(2\Omega) + \chi_Q(2\Omega)
\end{align}
with the Higgs (H) and quasiparticle (Q) contribution
\begin{subequations}
\begin{align}
    \chi_H(\omega) &= \chi_{\Delta A^2}(-\omega) \chi_{\Delta A^2}(\omega) H(\omega)\,,\\
    \chi_Q(\omega) &= \chi_{A^2 A^2}(\omega)\,.
\end{align}
\label{eq:k4_kernel}%
\end{subequations}
Comparing the response $j^{(3)}$
with the phenomenological Ginzburg-Landau model in Eq.~\eqref{eq:j_tdgl},
we can observe several differences which modify the response.
First, the Higgs propagator $H(\omega)$ is a more complex object
and does not have a simple resonance pole as we will see.
Second, light does not directly couple to the Higgs mode
but through the susceptibility $\chi_{\Delta A^2}(\omega)$.
Third, there is an additional quasiparticle response given by $\chi_{A^2A^2}(\omega)$.

In the following, let us disentangle these effects.
Evaluating the Matsubara sum
and rewriting the momentum sum as integral assuming $s$-wave symmetry,
the Higgs propagator can be analytically evaluated at $T=0$.
Concentrating on the pole structure one obtains
\begin{align}
    H(\omega) \propto \frac{1}{\sqrt{4\Delta^2 - \omega^2}}\,.
    \label{eq:Higgs_propagator2}
\end{align}
It does not have a simple pole
but a square root term in the denominator.
Transformed into time-domain, this leads to a power-law decay of the Higgs mode.
It can be understood as a decay into quasiparticles
as the Higgs mode energy overlaps with the quasiparticle continuum at $2\Delta$.
In addition to the obvious consequence of stronger damping,
it also affects the phase response.
The square root reduces the $\pi$ phase change at the resonance frequency
to $\pi/2$.
Thus, the driven amplitude oscillation only lags behind a quarter cycle
at high frequencies instead of being completely anti-phase
as found in the phenomenological model.

Next, we check how the electron bubbles $\chi_{\Delta A^2}(\omega)$
generating the light-Higgs coupling,
affects the phase response.
In the expression in Eq.~\eqref{eq:k4_kernel},
the term occurs twice evaluated at opposite frequency.
It can be written as its absolute squared value
\begin{align}
    \chi_{\Delta A^2}(\omega)\chi_{\Delta A^2}(-\omega) = |\chi_{\Delta A^2}(\omega)|^2
    \label{eq:dia_cancel}
\end{align}
and thus is not affecting the phase.

Finally, let us examine the quasiparticle response
which is actually known to be much larger than the Higgs response \cite{Cea2016}.
Evaluating the Matsubara sum of the respective susceptibility
and solving the momentum sum (see Appendix~\ref{app:higgs})
one obtains for the pole structure
\begin{align}
    \chi_{A^2 A^2}(\omega) &\propto
        \frac{1}{\sqrt{4\Delta^2 - w^2}}
        + \ldots
\end{align}
The quasiparticle response has the same square root pole
structure as the Higgs mode,
leading to the same $\pi/2$ phase change at the resonance frequency.

\begin{figure}[t]
    \centering
    \includegraphics[width=\columnwidth]{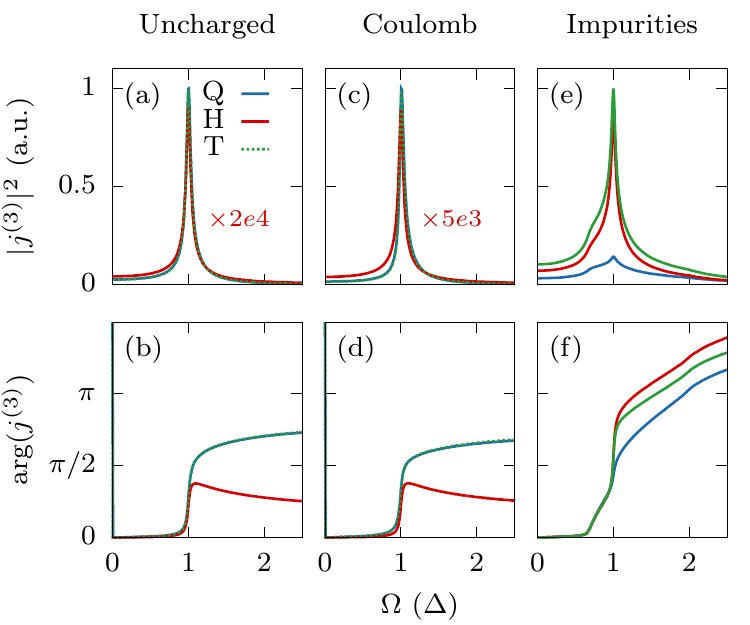}
    \caption{\label{fig:single_mode_micro}%
    Intensity (top row) and (normalized to zero) phase (bottom row)
    of THG response for Higgs (H), quasiparticles (Q) and total (T).
    (a),(b) Uncharged BCS model without Coulomb interaction in Eq.~\eqref{eq:k4_kernel}.
    The Higgs contribution is scaled by $2\cdot 10^4$ to be visible.
    (c),(d) BCS model including Coulomb interaction in Eq.~\eqref{eq:k4_coulomb}.
    The Higgs contribution is scaled by $5\cdot 10^3$ to be visible.
    (e),(f) BCS model with impurities using Mattis-Bardeen approach in Eq.~\eqref{eq:chi_dirty_limit}}
\end{figure}

In Fig.~\ref{fig:single_mode_micro}(a) and (b) the amplitude and phase
of the diamagnetic Higgs and quasiparticle response is shown
using $\epsilon_\vk = -2t(\cos k_x + \cos k_y) - \mu$,
$t=10$\,meV, $\mu=-10$\,meV, $\Delta=1$\,meV
and a residual broadening $\omega \rightarrow \omega + \im0.05$\,meV.
Hereby, the momentum sums are evaluated numerically on a 2d grid with $2000\times2000$ points
without approximation assuming linear polarized light in $x$-direction.
Confirming the analytic study,
we can see that phase shows a $\pi/2$ phase change
at the resonance frequency $2\Omega=2\Delta$.
Above the resonance, a drift is observable
to higher values for the quasiparticles and lower values for the Higgs mode.
As it has been emphasized in literature \cite{Cea2016},
the Higgs mode is much smaller in the clean-limit BCS theory.

\subsection{Influence of Coulomb interaction}
\label{sec:coulomb}
As a next step,
we discuss the influence of Coulomb interaction
given by an additional term in the Hamiltonian
\begin{align}
    H_c &= \frac 1 2 \sum_{\vk,\vk',\vq}\sum_{\sigma,\sigma'}
            V(\vq) c_{\vk+\vq,\sigma}^\dagger c_{\vk,\sigma}
                c_{\vk'-\vq,\sigma'}^\dagger c_{\vk',\sigma'}
\end{align}
where $V(\vq)$ is the Coulomb potential.
We follow \cite{Cea2016} and decouple the Coulomb interaction by means of an additional Hubbard-Stratonovich transformation
introducing the density field $\rho(\vq,\tau) = \rho_0 + \delta\rho(\vq,\tau)$
and allow amplitude and phase fluctuations
in the superconducting order parameter
$\Delta(\tau) = (\Delta + H(\tau))\e^{\im\theta(\tau)}$.
With this, one obtains for the fourth order action
\begin{align}
    & S^{(4)}(\delta\Delta,\theta,\delta\rho) = \frac 1 2 \frac 1 \beta
        \sum_{\im\omega_m} \Bigg[
            \phi^\top(-\im\omega_m)
            M(\im\omega_m)
            \phi(\im\omega_m)
        \notag\\& \qquad
            + \phi^\top(-\im\omega_m)
            b(\im\omega_m)
            + b^\top(-\im\omega_m)
            \phi(\im\omega_m)
    \notag\\&\qquad
    + \sum_{ijkl} A_{ij}^2(-\im\omega_m) A_{ij}^2(\im\omega_m)
        \chi_{A^2A^2}^{ijkl}(\im\omega_m)
    \Bigg]
    \label{eq:eff_action_coulomb}
\end{align}
with
\begin{subequations}
\begin{align}
    \phi^\top(\im\omega_m) &= \VVVT{\delta\Delta(\im\omega_m)}
        {\theta(\im\omega_m)}
        {\delta\rho(\im\omega_m)}\,,\\
    M &= \begin{pmatrix}
    H^{-1} &
    \frac{\im\omega_m}{2} \chi_{\Delta\rho} &
    \chi_{\Delta\rho} \\
    -\frac{\im\omega_m}{2} \chi_{\rho\Delta} &
    \frac{\omega_m^2}{4} \chi_{\rho\rho}&
    -\frac{\im\omega_m}{2} \chi_{\rho\rho}\\
    \chi_{\rho\Delta} &
    \frac{\im\omega_m}{2} \chi_{\rho\rho} &
    -\frac{1}{V(\vq)} + \chi_{\rho\rho}
    \end{pmatrix}\,,\\
    b(\im\omega_m) &= \VVV{
        \sum_{ij} A_{ij}^2(\im\omega_m) \chi_{\Delta A^2}^{ij}(\im\omega_m)
    }
    {
        -\im\omega_m
            \sum_{ij} A_{ij}^2(\im\omega_m) \chi_{\rho A^2}^{ij}(\im\omega_m)
    }
    {
            \sum_{ij} A_{ij}^2(\im\omega_m) \chi_{\rho A^2}^{ij}(\im\omega_m)
    }\,.
\end{align}
\end{subequations}
The susceptibilities are given in Appendix~\ref{app:coulomb}.
Integrating the fluctuations and using $1/V(\vq)\rightarrow 0$
for $\vq \rightarrow 0$
one obtains for the third-order current $j^{(3)} = \chi_H + \chi_Q$
\begin{subequations}
\begin{align}
    \chi_Q &=
        \chi_{A^2A^2} - \frac{|\chi_{A^2\rho}|^2}{\chi_{\rho\rho}}\,,\\
    \chi_H &=
        \frac{|\chi_{A^2\Delta} - \chi_{A^2\rho}\chi_{\Delta\rho}/\chi_{\rho\rho}|^2}
            {H^{-1} - |\chi_{\Delta_\rho}|^2/\chi_{\rho\rho}}\,.
\end{align}
\label{eq:k4_coulomb}%
\end{subequations}
The Coulomb interaction renormalizes the Higgs and quasiparticle response.
Yet, due to obtained structure, the phase signature is not changed
as the expressions in the nominator do not contribute due to the absolute square
and the $\chi_{\rho\rho}$ term has the same phase behavior as the unrenormalized propagators.
This can be seen in Fig.~\ref{fig:single_mode_micro}(c),(d),
where the respective expressions are numerically evaluated
with the same parameters of the previous section.
Except global scaling factors
and small deviations resulting from the $1/\chi_{\rho\rho}$ contribution
the result is basically unchanged with a phase change of $\pi/2$ at the resonance.

\subsection{Influence of impurities}
\label{sec:impurities}
Recently it was pointed out by several papers
\cite{Jujo2015,Murotani2019,Silaev2019,Tsuji2020,Seibold2021,Haenel2021}
that nonmagnetic impurities allow an additional paramagnetic coupling of light to the condensate.
This is shown diagrammatically in Fig.~\ref{fig:single_mode_diagram}(c) and (d)
where the light-coupling vertices are in the $\tau_0$-channel. While these diagrams vanish in the clean limit,
they have been shown to dominate the optical nonlinear response even for small
disorder. Here, we adopt the
Mattis-Bardeen approximation first applied to the nonlinear response in \cite{Murotani2019} and subsequently formulated in the effective action framework in \cite{Haenel2021}.

Following \cite{Haenel2021}, we implement a 3D continuum model,
where we can express the THG current within the
Mattis-Bardeen approximation as
\begin{align}
    j^{(3)}(3\Omega) &= \chi_H(2\Omega) + \chi_Q(2\Omega)
\end{align}
with the Higgs (H) and quasiparticle (Q) susceptibilities
\begin{subequations}
\begin{align}
    \chi_H(2\Omega) &= 2\chi_{AA\Delta}(2\Omega,-\Omega)
    \notag\\&\qquad \times
    \chi_{AA\Delta}(-2\Omega,-\Omega) H(2\Omega)\,,\\
    \chi_Q(2\Omega) &= \chi_{AAAA}(\Omega,2\Omega,-\Omega) \,.
\end{align}
\label{eq:chi_dirty_limit}%
\end{subequations}
The triangle and square bubbles are defined as
\begin{subequations}
\begin{align}
    &\chi_{AA\Delta}(\omega_m,\omega_l) = \frac 1 \beta
    \sum_{\omega_n}\sum_{\vk\vk'}
    |J_{\bm{kk'}}|^2
    \tr \Big[
    G_{0}(\omega_n+\omega_m, \bm{k})
    \notag\\&\qquad\times
    G_{0}(\omega_n+\omega_m+\omega_l, \bm{k'})
    G_{0}(\omega_n, \bm{k})
    \tau_1
    \Big]
    \,,\\
    &\chi_{AAAA}(\omega_m,\omega_l,\omega_p) = \frac 1 \beta \sum_{\omega_n}
        \sum_{\vk\vk'\vk''}
    |J_{\bm{kk'}}|^2
    |J_{\bm{kk''}}|^2
    \notag\\&\qquad\times
    \tr \Big[
        G_{0}(\omega_n,\bm{k})
        G_{0}(\omega_n+\omega_m,\bm{k'})
    \notag\\&\qquad\times
        G_{0}(\omega_n+\omega_m+\omega_l,\bm{k})
        G_{0}(\omega_n+\omega_m+\omega_l+\omega_p,\bm{k''})
    \Big]
\end{align}
\end{subequations}
and are shown in Fig.~\ref{fig:single_mode_diagram}(c) and (d).
The transition matrix element
$J_{\vk\vk'} = \braket{\vk|\frac{e\vec p}{m}|\vk'}$
is approximated by a Lorentzian distribution
\begin{align}
    \left|J_{\bm{kk'}}\right|^2
    \approx \frac{(e v_{F})^2}{3 N(0)}
    \frac{1}{\pi}\frac{\gamma}{\left(\epsilon_{\bm{k}}-\epsilon_{\bm{k}'}\right)^2
    + \gamma^2}
\end{align}
with impurity scattering rate $\gamma$, Fermi velocity $v_F$, and density of
states at Fermi surface $N(0)$.
We choose the parameters $\Delta=2$\,meV,
mass $m=0.78 m_e$ of the parabolic band dispersion,
$\epsilon_F=1$\,eV, and impurity scattering rate $\gamma/\Delta=10$.
We evaluate Matsubara sums analytically and
numerically compute the momentum integrals.
For further details about the calculation see \cite{Haenel2021}.
While the Mattis-Bardeen
approximation may not quield qualitatively accurate results in the nonlinear
response, it serves well to discuss qualitative differences of the phase
response compared to the clean limit.

The resulting amplitude and phase of the dirty superconductor are shown in Fig.~\ref{fig:single_mode_micro}(e) and (f).
We find a pronounced resonance peak at $2\Omega=2\Delta$. Here, the Higgs
contribution is no longer subdominant but instead gives the main contribution to
the THG signal.
The resonance peak is accompanied by a positive phase jump of roughly $\pi$
across the resonance. The detailed structure of this phase response as
well as the value of the phase jump show some weak dependence on material
parameters.

The more complex phase structure in the dirty limit can be understand
as follows: While the clean phase response is only given by the Higgs
propagator due to the cancellation of the phase of electronic susceptibilities
$\chi_{\Delta,A^2}$ in Eq.~\eqref{eq:dia_cancel},
the phase of the susceptibilities in the dirty case no longer cancels
but gives an additional additive contribution to the phase.
It is represented by the fermionic triangles
$\chi_{AA\Delta}(2\Omega,-\Omega) \chi_{AA\Delta}(-2\Omega,-\Omega)$
and shown in Fig.~\ref{fig:single_mode_diagram}(c).
Thus, the phase response in
the dirty limit is not only given by the Higgs propagator but
has an additional contribution from the electron-mediated
microscopic coupling of light to the Higgs mode.

\section{Phase response of two modes}
\label{sec:multiple-modes}
Now, we will consider systems which contain two modes
and study the interaction between these.
Again, we start by an analysis of the classic analogon
of two coupled oscillators to understand the fundamental properties
before proceeding to a microscopic model.

\subsection{Coupled oscillators}
\label{sec:coupled-ho}
If there are two modes in a system,
interference effects occur in the driven system
which can lead to the so-called antiresonance phenomenon.
The usual way to understand this effect is based on the assumption
that there are two modes in the system which are coupled
and only \emph{one} of these modes is externally driven.
For a particular driving frequency,
the external force on the driven mode
cancels exactly with the force induced by the other coupled mode
such that the amplitude of the oscillation of this mode vanishes
-- thus the name antiresonance.
Furthermore, the antiresonance
is accompanied by a negative phase jump of $\pi$,
therefore it goes in the opposite direction compared to a resonance.

The same phase signature can also be obtained when \emph{both} oscillators
are driven and the observed signal is comprised of the sum of both oscillation amplitudes.
Here, this effect is a trivial consequence of a destructive interference
and does not necessarily rely on a coupling between the modes.
An additional coupling between the modes
allows for a tuning of the antiresonance frequency.
We refer to this scenario as \emph{antiresonance behavior} as well.

To make this effect more clear,
let us first investigate again the classic model
where we consider now two coupled and driven oscillators
described by the following equations of motion
\begin{align}
    x_1''(t) + \omega_1^2 x_1(t) + \gamma_1 x_1'(t) + g x_2(t) &= F_1 \cos(\Omega t)\,,
    \notag\\
    x_2''(t) + \omega_2^2 x_2(t) + \gamma_2 x_2'(t) + g x_1(t) &= F_2 \cos(\Omega t)\,.
    \label{eq:coupled_oscillator}
\end{align}
There are two oscillators $x_{1}(t)$ and $x_{2}(t)$
with individual eigenfrequencies $\omega_i$, dampings $\gamma_i$ and
driving amplitudes $F_i$ but same driving frequencies $\Omega$.
The coupling between the modes is controlled by the constant $g$.
Using the complex variable method ansatz
\begin{align}
    x_i(t) = A_1\cos(\Omega t - \phi_i) = \Re \hat x_i(t)\,,
    \label{eq:coupled_modes_Ai}
\end{align}
where $\hat x_i(t) = \hat A_1(\Omega) \e^{\im\Omega t}$
with $\hat A_i(\Omega) = A_i(\Omega) \e^{-\im\phi_i(\Omega)}$,
we write the equations in matrix form
\begin{align}
    \MM{P_1^{-1}}{g}{g}{P_2^{-1}}
    \VV{\hat A_1}{\hat A_2} &=
    \VV{F_1}{F_2}\,,
    \label{eq:coupled_oscillator_M}
\end{align}
where we define the ``propagator'' of the oscillators as
$P_i^{-1} = -\Omega^2 + \omega_i^2 + \im\Omega\gamma_i$.
Inversion of the matrix leads to the solution
\begin{align}
    \VV{\hat A_1}{\hat A_2} &=
    \MM{\tilde P_1}{-gP_1\tilde P_2}{-gP_1\tilde P_2}{\tilde P_2}
    \VV{F_1}{F_2}
\end{align}
with the renormalized propagator $\tilde P_i = (P_i^{-1} - g^2 P_j)^{-1}$
where $i\neq j$.
We also consider the total response $x_T = x_1 + x_2$, where
\begin{align}
    x_T(t) = A_T \cos(\Omega t - \phi_T) = \Re \hat x_T(t)
    \label{eq:coupled_modes_AT}
\end{align}
with $\hat x_T(t) = \hat A_T(\Omega)\e^{\im\Omega t}$
and $\hat A_T = A_T(\Omega)\e^{-\phi_T(\Omega)}$.
One obtains for the complex amplitudes
\begin{subequations}
\begin{align}
    \hat A_1 &= \tilde P_1 F_1 - gP_1\tilde P_2 F_2\,, \\
    \hat A_2 &= \tilde P_2 F_2 - gP_1\tilde P_2 F_1\,, \\
    \hat A_T &= \tilde P_1 F_1 + \tilde P_2 F_2 - g P_1\tilde P_2 (F_1+F_2)\,.
\end{align}
\label{eq:coupled_modes_A_phi}%
\end{subequations}
\begin{figure}[t]
    \centering
    \includegraphics[width=\columnwidth]{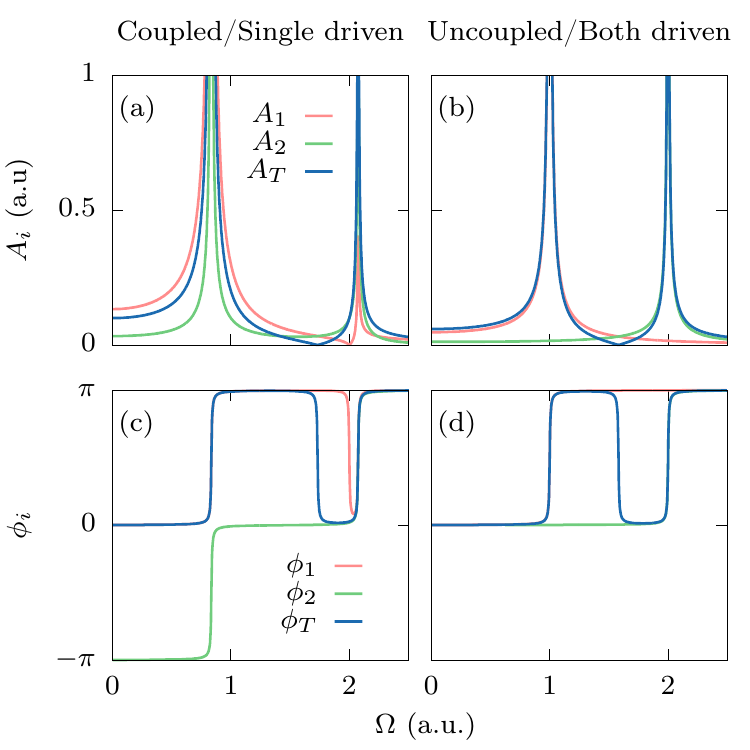}
    \caption{\label{fig:coupled_mode}%
    (a),(b) Amplitude and (c),(d) phase of two (coupled) modes
    as defined in Eq.~\eqref{eq:coupled_modes_Ai} and Eq.~\eqref{eq:coupled_modes_AT}.
    For the left column (a) and (c), the two oscillators are coupled with $g=1$
    but only the first oscillator is driven with $F_1 = 1$ and $F_2 = 0$.
    For the right column (b) and (d), the two oscillators are uncoupled
    with $g=0$  but both oscillators are driven with $F_1 = F_2 = 1$.
    The frequencies are $\omega_1 = 1$, $\omega_2 = 2$
    and the dampings $\gamma_1 = \gamma_2 = 0.01$.}
\end{figure}%
In Fig.~\ref{fig:coupled_mode} we show a numerical evaluation
of the individual and total amplitudes $A_i$ and phases $\phi_i$
for two distinct cases (see Appendix~\ref{app:coupled_oscillators} for the exact expressions).
In the first column, the two oscillators are coupled, i.e. $g\neq 0$,
but only the first oscillator is driven $F_2 =0$.
In the second column, the two oscillators are uncoupled, i.e. $g=0$,
but both oscillators are driven $F_i \neq 0$.

The first scenario (left column) corresponds
to the usual definition of the antiresonance,
namely a destructive interference between the driving force and
the force from the second oscillator due to the coupling.
The dip between the two resonance peaks
and the negative $\pi$ phase change, is clearly visible for the first oscillator (red curve).
The energy of the antiresonance $\omega_A$ is determined by $\tilde P_1 = 0$,
which leads to $\omega_A = \omega_2$,
i.e. the antiresonance occurs at the energy of the other undriven mode.
The total response $A_T$ and $\phi_T$ (blue curve)
also shows this behavior
resulting from the antiresonance of the first oscillator.
Yet, the energy of the antiresonance is shifted
as a result of the second superposition scenario.

We can further see that the finite coupling shifts the resonances frequency
with respect to the uncoupled eigenfrequencies $\omega_i$.
The resonance frequency for the lower modes is decreases,
while the resonance frequency  of the higher mode is increased.
The energies are given by the poles of the renormalized propagators
\begin{align}
    \tilde \omega_i = \frac{1}{\sqrt{2}}\sqrt{\omega_1^2 + \omega_2^2 \pm \sqrt{(\omega_1^2 -\omega_2^2)^2 + 4g^2}}\,.
\end{align}
For the shown parameters, this results in a resonance peak below $\omega_1 = 1$
and a resonance peak above $\omega_2 = 2$.

The total response (blue curve) of the second scenario (right column) shows a very similar behavior,
namely a dip in between the two resonance peaks
and a negative $\pi$ phase change.
However, in this case the negative phase jump does not result
from an individual oscillator,
both individual oscillators (red and green curve) do not show this behavior.
It rather results from the superposition of the two oscillations
where the sum of both cancel out
in an intermediate position between the resonances.
This energy is determined by
$P_1 + P_2 = 0$ leading to $\omega_A = \frac{1}{\sqrt{2}}\sqrt{\omega_1^2 + \omega_2^2}$.
As the sum of both undergoes a sign change from negative to positive
a negative $\pi$ phase change occurs naturally.
The resonance frequencies in this scenario are not changed
and still occur at $\omega_i$.

To summarize, the antiresonance behavior of the total response of two oscillators
can have different origins.
It can be either controlled by the coupling between the oscillators
or the interference if both modes are driven.

\begin{figure}[t]
    \centering
    \includegraphics[width=\columnwidth]{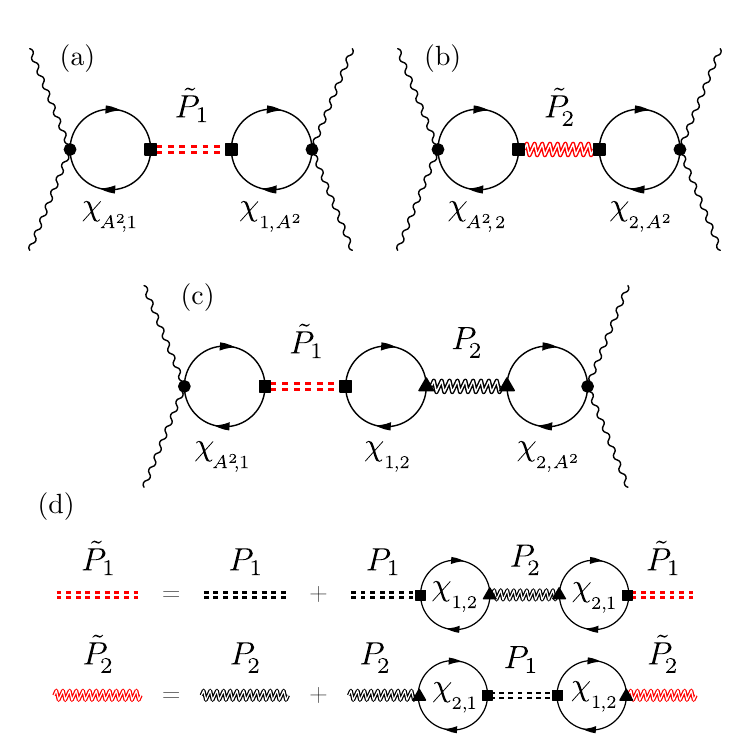}
    \caption{\label{fig:coupled_mode_diagram}%
    Diagrammatic representation of effective action for a system with Higgs and another mode
    assuming diamagnetic coupling to light.
    (a),(b) Excitation of renormalized mode 1 or 2
    (c) Mixed contribution term, where light couples to both modes and the modes to each other.
    (d) Renormalization of both modes as RPA series due to interaction with each other.
    The wiggly lines represent the vector potential $A$,
    the solid lines the BCS Green's function $G_0$,
    the double dashed line the Higgs propagator $H$
    and the double zigzag line the propagator of another mode.
    Red lines represent the renormalized propagators.
    The filled square vertex corresponds to $f_\vk \tau_1$,
    the filled circle vertex to $\partial_{ij}^2\epsilon_\vk \tau_3$
    and the filled triangle vertex represent the interaction with the other mode.
    }
\end{figure}%

\subsection{Microscopic theory}
\label{sec:micr-coupled}
Let us now investigate whether
this behavior is observable in a microscopic model as well.
For now, we will make some general arguments
assuming that there are two modes $1$ and $2$ in the system,
for example the Higgs mode and a second collective mode.
In Sec.~\ref{sec:higgs-cdw} and Sec.~\ref{sec:higgs-bs}
we will consider specific examples.

Taking into account the general form of the effective action
and the analysis of the classical oscillator system,
we are anticipating the results of the next sections
and postulate the general structure of the response.
The fourth order effective action for two modes reads
\begin{align}
    S^{(4)} &= \frac 1 2 \frac 1 \beta \sum_{\im\omega_m}
    b^\top(-\im\omega_m) M^{-1}(\im\omega_m)b(\im\omega_m)
    \notag\\
    &= \frac 1 2 \frac 1 \beta \sum_{\im\omega_m} K^{(4)}(\im\omega_m) A^2(-\im\omega_m) A^2(\im\omega_m)
    \label{eq:S4_two_modes_matrix}
\end{align}
where
\begin{subequations}
\begin{align}
M^{-1} &= \MM{\tilde P_1}{\chi_{1,2}P_1 \tilde P_2}
    {\chi_{2,1}P_1 \tilde P_2}{\tilde P_2}\,,\\
b &= \VV{\chi_{1,A^2}}{\chi_{2,A^2}} A^2\,.
\end{align}
\end{subequations}
Hereby, $P_i$ stands for the propagator of mode $i$,
$\chi_{i,j}$ for different coupling susceptibilities
and $A$ the vector potential,
where the polarization indices are not included.
The tilde denotes a renormalization due to the other mode.
This can be understood as an RPA series renormalization of the propagators
shown in Fig.~\ref{fig:coupled_mode_diagram}(d) and expressed as
\begin{subequations}
\begin{align}
    \tilde P_{1} &= P_{1} + |\chi_{1,2}|^2 P_{1} P_{2} \tilde P_{1}\,, \\
    \tilde P_{2} &= P_{2} + |\chi_{1,2}|^2 P_{1} P_{2} \tilde P_{2}
\end{align}
\end{subequations}
which leads to
\begin{subequations}
\begin{align}
    \tilde P_{1} &= \frac{1}{P_{1}^{-1}-|\chi_{1,2}|^2 P_{2}}\,,\\
    \tilde P_{2} &= \frac{1}{P_{2}^{-1}-|\chi_{1,2}|^2 P_{1}}\,.
\end{align}
\end{subequations}
The fourth-order kernel $K^{(4)}$ explicitly reads
\begin{align}
    K^{(4)} &= \chi_1 + \chi_2 + \chi_{12}
\end{align}
with
\begin{subequations}
\begin{align}
    \chi_1 &= |\chi_{1,A^2}|^2 \, \tilde P_{1}\,, \label{eq:chi_M1}\\
    \chi_2 &= |\chi_{2,A^2}|^2 \, \tilde P_{2}\,, \label{eq:chi_M2}\\
    \chi_{12} &= P_{1} \tilde P_2(
        \chi_{A^2,1} \, \chi_{1,2} \, \chi_{2,A^2}
        +\chi_{A^2,2}  \, \chi_{2,1} \, \chi_{1,A^2}
        )\,. \label{eq:chi_M1_M2}
\end{align}
\end{subequations}
These terms are diagrammatically
shown in Fig.~\ref{fig:coupled_mode_diagram}(a)-(c)
and can be understood in the following way.
First of all, both modes may couple individually to light represented by
Eq.~\eqref{eq:chi_M1} and Eq.~\eqref{eq:chi_M2}.
If there is a coupling between the modes,
a mixed term Eq.~\eqref{eq:chi_M1_M2} occurs,
where light couples to both modes and the modes to each other.

A comparison with the classic coupled oscillator model
of the previous sections in Eq.~\eqref{eq:coupled_modes_A_phi}
reveals the exact same structure except that there,
all susceptibilites are constant without frequency dependence.

After these general remarks,
let us now consider specific examples
of two microscopically coupled modes in the next sections.

\section{Higgs and Charge density wave}
\label{sec:higgs-cdw}
\begin{figure}[t]
    \centering
    \includegraphics[width=\columnwidth]{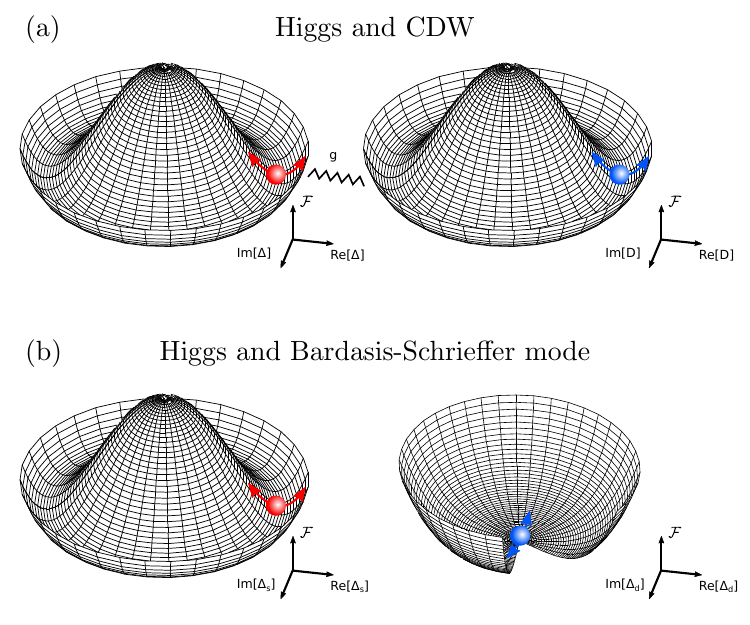}
    \caption{\label{fig:free_energy}%
    Collective modes in the picture of the free energy.
    (a) Higgs and CDW mode. The Higgs mode is the amplitude fluctuation of the superconducting order parameter
    and the CDW mode the amplitude fluctuation of the CDW order parameter corresponding to the renormalized CDW phonon.
    (b) Higgs and Bardasis-Schrieffer mode. The Bardasis-Schrieffer mode is the amplitude oscillation of the subleading pairing channel orthogonal to the amplitude (Higgs) oscillation of the dominant pairing channel.}
\end{figure}

As a first example of two coupled collective modes,
we will consider a coexisting
superconducting and charge density wave (CDW) system.
The amplitude modes are schematically shown in Fig.~\ref{fig:free_energy}(a)
in the picture of the free energy.
An example for such a scenario is NbSe$_2$,
where the coupling of the Higgs mode to a CDW phonon was
observed in Raman response \cite{Sooryakumar1980,Measson2014}
and theoretically investigated by several authors \cite{Littlewood1981,Browne1983,Cea2014}.
Another relevant system are cuprates,
where superconductivity and fluctuating charge order
has been reported in the underdoped regime \cite{Torchinsky2013,Hinton2013}.
This scenario might be a possible explanation of the antiresonance behavior
observed in recent THG experiments \cite{Chu2020}.

To model the system, we follow \cite{Cea2014}
and start from the BCS Hamiltonian in Eq.~\eqref{eq:bcs_hamiltonian}
where we add a phonon of momentum $\vQ$ responsible for creating the charge order
and a coupling to electrons with strength $g$.
The Hamiltonian is given by
\begin{align}
    H &= H_{\mathrm{BCS}} + H_{\mathrm{CDW}}
    \label{eq:H_cdw}
\end{align}
with
\begin{align}
    H_{\mathrm{CDW}} &= \sum_{\mathclap{\vq=\pm\vQ}}
            \omega_\vq  b_\vq^\dagger b_{\vq}
        + g \sum_{\mathclap{\vk,\vq=\pm\vQ,\sigma}} g_{\vk}
            c_{\vk+\vq,\sigma}^\dagger c_{\vk\sigma}(b_\vq + b_{-\vq}^\dagger)\,.
\end{align}
Hereby, $b_\vq^\dagger$ or $b_\vq$
are the phonon creation or annihilation operators
and $\omega_Q$ the energy of the CDW phonon.
The electron phonon coupling is controlled by $g\cdot g_\vk$
with strength $g$ and momentum dependence $g_\vk$.

To simplify the calculation, we will make the following assumptions.
We consider a 2d square lattice with a tight-binding dispersion
and nearest-neighbor hopping $t$ at half-filling,
namely $\epsilon_\vk = -2t(\cos k_x + \cos k_y)$.
As it was shown in \cite{Cea2014},
a finite chemical potential leads to a qualitative similar result.
Choosing $\vQ = \VVT{\pi}{\pi}$ we have perfect nesting
and a commensurate CDW with $\vk + 2\vQ \hat = \vk$
where
$\epsilon_{\vk+2\vQ} = \epsilon_{\vk}$ and $\epsilon_{\vk+\vQ} = -\epsilon_\vk$.
We assume an $s$-wave superconductor with $f_\vk = 1$
and an anisotropic $s$-wave CDW with $g_\vk = |\cos k_x - \cos k_y|$.

We start from the action of the system,
where we introduce a CDW order parameter $D_\vk = D g_\vk$
with $D = g\braket{b_\vQ + b_{-\vQ}^\dagger}$ and the
superconducting order parameter $\Delta$
using a Hubbard-Stratonovich transformation.
Details of the calculation can be found in Appendix~\ref{app:higgs-cdw}.
Please note that we neglect here the Coulomb interaction and phase fluctuations
as we have shown in Sec.~\ref{sec:coulomb}
that they do not affect the phase signature.
Furthermore, in the half-filled case, as considered here,
its influence vanishes completely
as the system has perfect particle-hole symmetry \cite{Cea2014}.

After integration of the fermions and expansion of the action at Gaussian level
for amplitude fluctuations $\Delta(t) = \Delta + \delta\Delta(t)$
and $D(t) = D + \delta D(t)$,
the effective fourth order action reads
\begin{align}
    & S^{(4)} = \frac 1 2\frac 1 \beta \sum_{\im\omega_m} \Bigg(
    \phi^T(-\im\omega_m) M(\im\omega_m) \phi(\im\omega_m)
    \notag\\&\qquad
    + \phi^T(-\im\omega_m)b(\im\omega_m) + b^\top(-\im\omega_m)\phi(\im\omega_m)
    \Bigg)
    \label{eq:effective_action_fluctuations}
\end{align}
with
\begin{subequations}
\begin{align}
    \phi^T(-\im\omega_m) &= \VVT{\delta\Delta(-\im\omega_m)}
        {\delta D(-\im\omega_m)}\,,\\
    M(\im\omega_m) &= \MM{H^{-1}(\im\omega_m)}
        {\chi_{\Delta D}(\im\omega_m)}
        {\chi_{D\Delta}(\im\omega_m)}
        {-\frac{1}{g^2}P^{-1}(\im\omega_m)}\,,\\
    b(\im\omega_m) &= -\sum_{ij} A_{ij}^2(\im\omega_m)
        \VV{\chi_{\Delta A^2}^{ij}(\im\omega_m)}{\chi_{D A^2}^{ij}(\im\omega_m)}\,,
\end{align}
\end{subequations}
where $H$ is the Higgs propagator and $P$ the renormalized Phonon propagator.
It is defined as
\begin{align}
    P^{-1}(\im\omega_m) &= P_0^{-1}(\im\omega_m) - g^2\chi_{DD}(\im\omega_m)
\end{align}
with the bare phonon propagator $P_0 = -2\omega_Q/(\omega_Q^2 - (\im\omega_m)^2)$
and $\chi_{DD}$ the susceptibility describing the influence of the CDW.
The susceptibility on the off-diagonal in $M$
leads to a coupling between Higgs and CDW
and the expression is equivalent
to the coupled oscillator system Eq.~\eqref{eq:coupled_oscillator_M} in the previous section.
An integration of the amplitude fluctuations finally leads to
\begin{align}
    & S^{(4)} = \frac 1 2 \frac 1 \beta \sum_{\im\omega_m}
        b^\top(-\im\omega_m) M^{-1}(\im\omega_m) b(\im\omega_m)
    \label{eq:eff_action_cdw}
\end{align}
with
\begin{align}
    M^{-1} &= \MM{\tilde H}
        {g^2\chi_{\Delta D}P\tilde H}
        {g^2\chi_{D \Delta}\tilde PH}
        {-g^2 \tilde P}
\end{align}
where we identify the renormalized Higgs and phonon propagator
as shown diagrammatically in Fig.~\ref{fig:coupled_mode_diagram}(d)
\begin{subequations}
\begin{align}
    \tilde H &= \frac{1}{H^{-1}
        + g^2 \chi_{\Delta D}\chi_{D\Delta}
        P}\,,\\
    \tilde P &= \frac{1}{P^{-1}
        + g^2 \chi_{\Delta D}\chi_{D\Delta}
        H}\,.
\end{align}
\end{subequations}
The expression of the action follows exactly the general structure
as shown in the previous section and the resulting diagrams
are the ones shown in Fig.~\ref{fig:coupled_mode_diagram}(a)-(c).

Let us analyze the result in more detail.
First, we evaluate the expression for the bare Higgs propagator.
One finds
\begin{align}
    H(\omega) &\propto
            \frac{\sqrt{4\Delta^2+4D^2-\omega^2}}
                {4\Delta^2 - \omega^2}\,.
\end{align}
For $D = 0$, we would retain expression Eq.~\eqref{eq:Higgs_propagator2}.
However, for finite CDW gap $D$, the Higgs mode energy $2\Delta$ no longer
coincidences with the quasiparticle excitation gap $\Delta + D$
and, as already pointed out by \cite{Cea2014}, the Higgs mode becomes stable
as its energy is below the gap.
This has consequences for the pole structure,
as the Higgs mode now has a simple pole without square root
such that we can expect a $\pi$ phase shift
when varying the driving frequency $\omega$
from below to above the Higgs mode energy.

The CDW phonon propagator can be evaluated and reads
\begin{align}
    P(\im\omega_m) &= -\frac{2\omega_Q}{\Omega_Q^2 - (\im\omega_m)^2}
\end{align}
with
\begin{align}
    \Omega_Q^2 &= 4g^2\omega_Q \sum_{\vk} g_\vk^2
            \frac{4D_\vk^2 - (\im\omega_m)^2}
        {E_\vk(4E_\vk^2 - (\im\omega_m)^2)}
        \tanh(\beta E_\vk/2)\,.
\end{align}
The phonon propagator has a simple pole leading to a phase change of $\pi$.

Assuming linear polarized light in $x$-direction as in Sec.~\eqref{sec:bcs},
we can write the action as
\begin{align}
    S^{(4)} &= \frac 1 2 \int \mathrm d\omega\, K^{(4)}(\omega) A^2(-\omega) A^2(\omega)\,,
\end{align}
where the kernel is given by $K^{(4)} = \chi_H + \chi_P + \chi_M$ with
the Higgs (H), phonon (P) and mixed (M) contributions
\begin{subequations}
\begin{align}
    \chi_H &=
        |\chi_{A^2\Delta}|^2 \tilde H \,,\\
    \chi_P &= - g^2 |\chi_{A^2 D}|^2 \tilde P \,,\\
    \chi_M &= g^2 \tilde H P
            (\chi_{A^2 \Delta} \chi_{A^2D} \chi_{\Delta D}
            + \chi_{\Delta A^2} \chi_{D A^2} \chi_{D \Delta})\,.
\end{align}
\end{subequations}
It has the same structure as Eq.~\eqref{eq:chi_M1_M2}
introduced as the general response for coupled modes.
With the insight of the previous sections,
we can expect an antiresonance behavior
with a negative phase change of $\pi$
in between the two resonances
where a phase change of positive $\pi$ should occur.

\begin{figure}[t]
    \centering
    \includegraphics[width=\columnwidth]{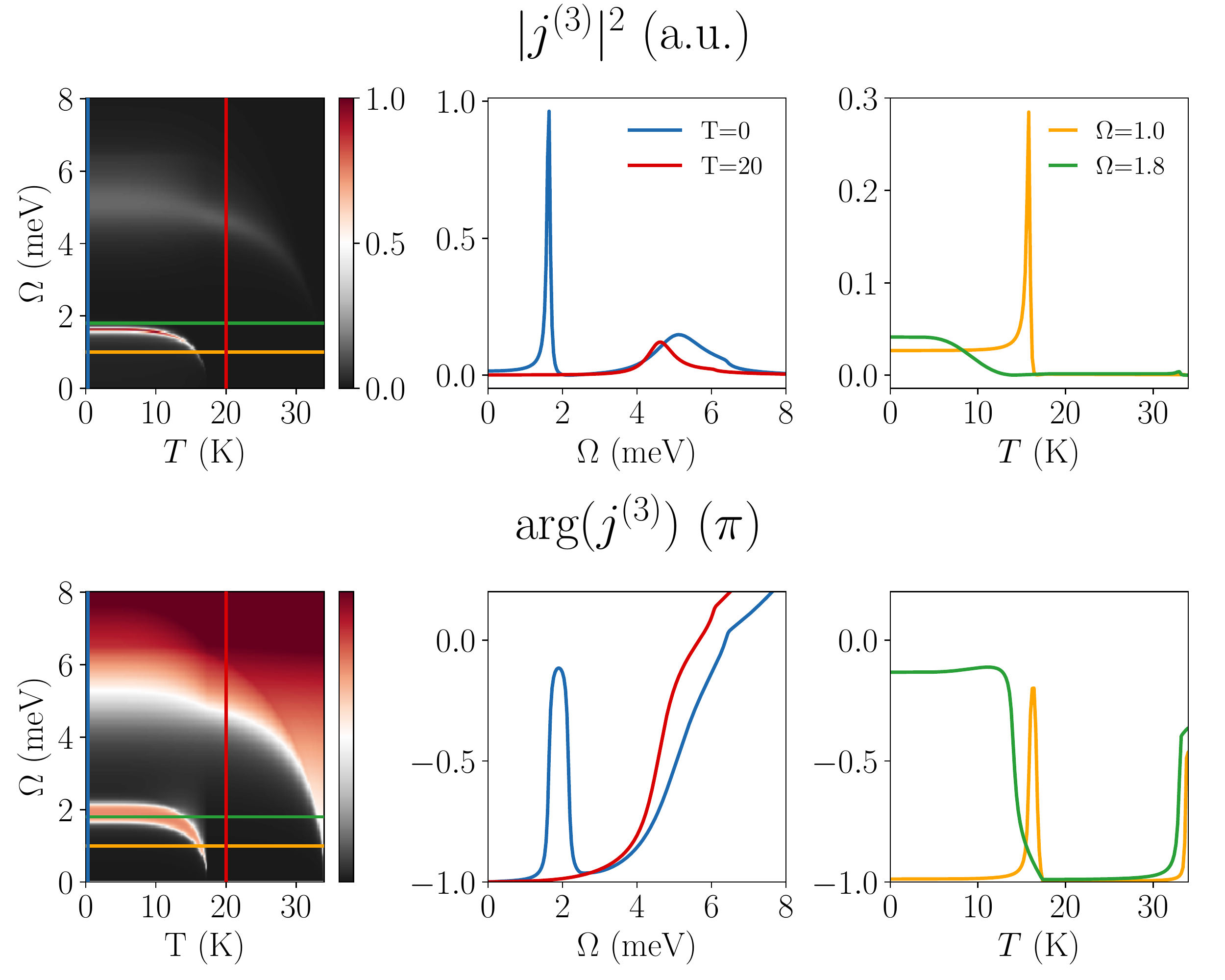}
    \caption{\label{fig:higgs_cdw}%
    Intensity (top row) and phase (bottom row) of THG signal in the coexisting
    superconducting and CDW state as function of temperature and frequency.
    The first column shows the full temperature and frequency dependence.
    The second column shows the frequency dependence for selected temperatures (vertical cuts). The third column shows the temperature dependence for selected frequencies (horizontal cuts).}
\end{figure}

To confirm, we calculate numerically the total THG response
as function of frequency and temperature.
The temperature dependence is necessary to compare with experimental results
where only the temperature can be varied for fixed driving frequency.
The calculation for a set of parameters $\Delta = 2.5$\,meV, $\omega_0=15$\,meV,
$D=3$\,meV, $t=10$\,meV, $\omega \rightarrow \omega+\im0.1$\,meV
is evaluated on a 2d grid with $2000\times2000$ points and
shown in Fig.~\ref{fig:higgs_cdw}.
Hereby, the CDW phonon energy is above the Higgs mode energy.
The top row shows the THG intensity,
and the bottom row the THG phase.
The first column is a 2d plot of the THG signal as function of frequency
and temperature.
Thus, vertical cuts in this plot, shown in the second column,
correspond to varying the frequency for fixed temperature
and horizontal cuts, shown in the third column,
correspond to varying the temperature for fixed frequency.

The result fulfills our expectation of the previous general analysis.
Looking at the 2d plot in the first column,
we can see the Higgs mode as a sharp resonance peak
which follows the temperature dependence of the gap.
However, due to the coupling to the CDW,
the energy of the Higgs mode is renormalized and shifted to lower frequencies.
This resonance peak is accompanied by a positive phase jump of $\pi$
as the Higgs mode is a stable mode below the total gap as discussed before.
At a slightly higher energy,
we observe an antiresonance behavior with a dip in the amplitude
and a negative phase jump of $\pi$.
At higher energy, we observe a second resonance peak
at the renormalized CDW phonon frequency
with an associated positive phase jump of $\pi$.
This resonance peak follows the temperature dependence of the CDW gap.
Please note, due to the residual broadening,
the positive phase change at the Higgs mode and the negative change at the antiresonance
is slightly lower than $\pi$.

The temperature dependence of both modes are similar
and follow roughly a quarter circle
as can be seen in the left column in Fig.~\ref{fig:higgs_cdw}.
Thus, vertical cuts along the frequency and horizontal cuts along the temperature
reveal, in principle, the same information.
Resonance peaks and phase changes
are visible in both signals.
Yet, obtaining single cuts at unfavorable positions,
e.g. in between the modes,
or limited variation range of parameters
might miss several features.
Experiments with a large variation of temperature and frequency
are therefore necessary to reveal the full information.

\section{Higgs and Bardasis-Schrieffer mode}
\label{sec:higgs-bs}
As a second example,
we consider a superconducting system
where the ground state is dominated by one symmetry,
yet fluctuations in a subleading pairing channel are allowed.
These fluctuations are known as Bardasis-Schriefer modes
\cite{Bardasis1961,Sun2020}
and might exist for example in iron-based superconductors
\cite{Scalapino2009,Maiti2015,Maiti2016},
where multiple pairing instabilities occur on different bands.
While most studies investigated the signature of a Bardasis-Schrieffer mode in Raman spectra,
recent work has shown
that such modes can also be excited with THz light \cite{Mueller2019}.
This mode is schematically shown in Fig.~\ref{fig:free_energy}(b)
and will be discussed in the following.

Here, we will consider an $s$-wave ground state and allow fluctuations in the $d$-wave channel.
We use the BCS Hamiltonian in Eq.~\eqref{eq:bcs_hamiltonian} with a sum of separable
pairing interactions
\begin{align}
    V_{\vk,\vk'} = V_s f_\vk^s f_{\vk'}^s + V_d f_\vk^d f_{\vk'}^d
    \label{eq:bs_Vkk}
\end{align}
with the (anisotropic) $s$-wave symmetry $f_\vk^s = (\cos k_x + \cos k_y)/2$ and
$d$-wave symmetry $f_\vk^d = (\cos k_x - \cos k_y)/2$.
Choosing $\epsilon_\vk = -2t(\cos k_x + \cos k_y) - \mu$
with $t=6$\,meV, we solve the gap equations
\begin{align}
    \Delta_i &= V_i \sum_\vk f_\vk^i \frac{\Delta_\vk}{2E_\vk}
\end{align}
with $\Delta_\vk = \sum_i \Delta_i f_\vk^i$ for $i=s,d$
for varying Fermi level $\mu$ and symmetry ratio $V_d/V_s$.
This phase diagram is shown in the Appendix in Fig.~\ref{fig:phase_diagram_bs}.
In the following, we choose parameters $\mu = -12$\,meV, $V_d = V_s$ and $\Delta=2$\,meV,
where the ground state is $s$-wave but still close to the $d$-wave transition.
The residual broadening is $\omega \rightarrow \omega + \im0.05$\,meV.
Please note, for the chosen parameters, i.e. the anisotropic $s$-wave and the energy dispersion,
the maximum of the gap at the Fermi level is $\Delta_{\text{max}} \approx 1$\,meV,
such that the Higgs mode energy is at $\omega_H \approx 2\Delta_{\text{max}}$
and not at $\omega_H = 2\Delta$.

If the system is excited,
we allow fluctuations in both symmetry channels
\begin{subequations}
\begin{align}
    \Delta_s(t) &= \Delta_s + \delta\Delta_s(t)\,,\\
    \Delta_d(t) &= \im \delta\Delta_d(t)\,.
\end{align}
\end{subequations}
Hereby, $\delta\Delta_s(t)$ corresponds to the usual Higgs mode
of the dominant symmetry channel,
while $\delta\Delta_d(t)$ are amplitude fluctuations
of the subleading channel orthogonal, i.e. in the imaginary axis direction.
This is the Bardasis-Schrieffer mode.
As shown in \cite{Sun2020}, subleading fluctuations
in the parallel or real channel
do not lead to a finite energy mode.

After integrating out the fermions (see Appendix~\ref{app:higgs-bs}),
we obtain the same structure of the effective action
as Eq.~\eqref{eq:effective_action_fluctuations}
with
\begin{subequations}
\begin{align}
    \phi^T(-\im\omega_m) &= \VVT{\delta\Delta_s(-\im\omega_m)}{\delta\Delta_d(-\im\omega_m)}\,,\\
    M(\im\omega_m) &= \MM{H^{-1}(\im\omega_m)}
        {-\chi_{\Delta B}(\im\omega_m)}
        {-\chi_{B\Delta}(\im\omega_m)}
        {B^{-1}(\im\omega_m)}\,,\\
    b(\im\omega_m) &= \sum_{ij} A_{ij}^2(\im\omega_m)
    \VV{-\chi_{\Delta A^2}^{ij}(\im\omega_m)}
    {\chi_{BA^2}^{ij}(\im\omega_m)}
    \label{eq:bs_S4}
\end{align}
\end{subequations}
where $H(\im\omega_m)$ is the usual Higgs propagator
and $B(\im\omega_m)$ the Bardasis-Schrieffer propagator defined as
\begin{align}
    B^{-1}(\im\omega_m) &= \frac{2}{V_d} + \chi_{BB}(\im\omega_m)\,.
\end{align}
The susceptibilities are defined as
\begin{subequations}
\begin{align}
    \chi_{BB}(\im\omega_m) &= \sum_\vk (f_\vk^d)^2 X_{22}(\vk,\im\omega_m)\,,\\
    \chi_{\Delta B}(\im\omega_m) &=
        \sum_\vk f_\vk^s f_\vk^d X_{12}(\vk,\im\omega_m)\,,\\
    \chi_{B A^2}^{ij}(\im\omega_m) &=
        \sum_\vk \frac 1 2 f_\vk^d \partial_{ij}^2\epsilon_\vk X_{23}(\vk,\im\omega_m)\,.
\end{align}
\end{subequations}
with $X_{\alpha\beta}$ defined in Eq.~\eqref{eq:XX}.
In analogy to the previous sections, the fluctuations are integrated out
which leads to
\begin{align}
S^{(4)} &= \frac 1 2 \frac 1 \beta \sum_{\im\omega_m}\Big(
b^\top(-\im\omega_m)M^{-1}(\im\omega_m)b(\im\omega_m)
\notag\\&\qquad
+ \sum_{ijkl} A_{ij}^2(-\im\omega_m)A_{kl}^2(\im\omega_m)\chi^{ijkl}_{A^2A^2}(\im\omega_m)
\Big)
\end{align}
with
\begin{align}
    M^{-1} &= \MM{\tilde H}
    {\chi_{\Delta B} \tilde H B}
    {\chi_{B \Delta} \tilde H B}
    {\tilde B}
\end{align}
and the renormalized propagators
\begin{subequations}
\begin{align}
    \tilde H &= \frac{1}{H^{-1}-\chi_{\Delta B}\chi_{B\Delta} B}\,, \\
    \tilde B &= \frac{1}{B^{-1}-\chi_{\Delta B}\chi_{B\Delta} H}\,.
\end{align}
\end{subequations}
For monochromatic, linear polarized light with polarization angle $\theta$,
the THG current parallel to the light polarization is given by
(see Appendix~\ref{app:higgs-bs})
\begin{align}
    j_{\parallel}^{(3)} &\propto (\cos^4\theta + \sin^4\theta) K_{xx}(2\Omega)
    \notag\\&\qquad
        + 2\sin^2\theta\cos^2\theta K_{xy}(2\Omega)
\end{align}
with the kernel $K^{(4)}_{ij} = \chi_H + \chi_B + \chi_M + \chi_Q$
and the susceptibilities for the Higgs (H), Bardasis-Schrieffer (B), mixed (M) and quasiparticle (Q) contribution
\begin{subequations}
\begin{align}
    \chi_H &= \chi_{\Delta A^2}^{ii}(-\omega)
    \chi_{\Delta A^2}^{jj}(\omega)
    \tilde H(\omega)\,,\\
    \chi_B &= \chi_{B A^2}^{ii}(-\omega)
    \chi_{B A^2}^{jj}(\omega)
    \tilde B(\omega)\,,\\
    \chi_M &=
    - \tilde H(\omega)
    B(\omega)\Big[
    \chi_{A^2 \Delta}^{ii}(\omega)
    \chi_{\Delta B}(\omega)
    \chi_{B A^2}^{jj}(\omega)
    \notag\\&\qquad
    +
    \chi_{A^2 B}^{ii}(\omega)
    \chi_{B \Delta}(\omega)
    \chi_{\Delta A^2}^{jj}(\omega)
    \Big]
    \,,\\
    \chi_Q &= \chi_{A^2A^2}^{iijj}(\omega)\,.
\end{align}
\label{eq:bs-suscept}%
\end{subequations}
The response has again the same structure of coupled modes as discussed before.
\begin{figure}[t]
    \centering
    \includegraphics[width=\columnwidth]{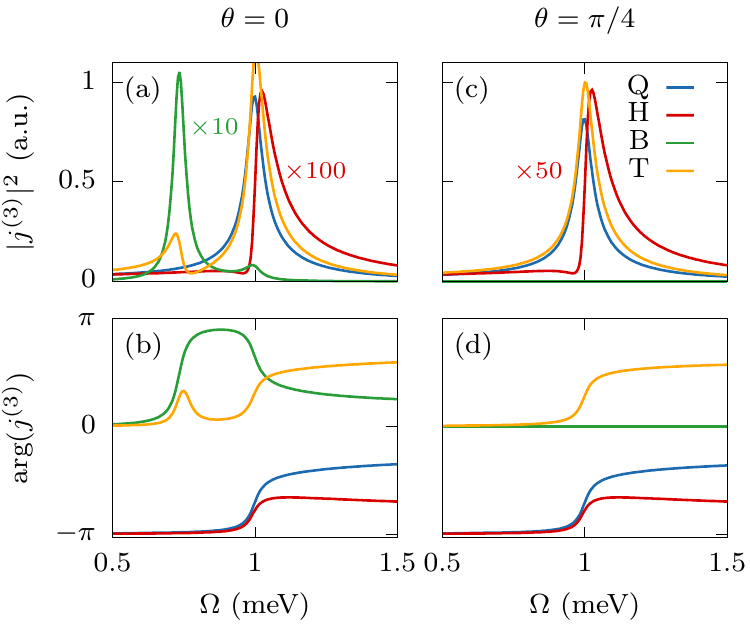}
    \caption{\label{fig:higgs_bs_contributions}%
    THG intensity (top row) and phase (bottom row)
    for Higgs and Bardasis-Schrieffer mode system
    for two light polarizations $\theta = 0$ (left column) and $\theta = \pi/4$ (right column).
    The individual contributions are shown separately as quasiparticles (Q),
    Higgs (H), Bardasis-Schrieffer mode (B), and total response (T).
    The Higgs and Bardasis-Schrieffer modes are scaled to be visible.}
\end{figure}%

To get a first insight into the Bardasis-Schrieffer mode,
we evaluate the expression for the propagator analytically for $T=0$
and assuming a constant density of states at the Fermi level.
One obtains for continuum implementation (see Appendix~\ref{app:higgs-bs})
\begin{align}
    B(\omega) &= \frac{\sqrt{4\Delta^2-\omega^2}}
    {\left(
    \frac{2}{V_d} - \frac{2}{V_s}\right)\sqrt{4\Delta^2 - \omega^2}
    - 2 \omega \lambda \sin^{-1}\left(\frac{\omega}{2\Delta}\right)
    }\,.
\end{align}
Using these simplification,
one can see that for $V_d \rightarrow V_s$
the pole of the propagator shifts to zero,
while for $V_d \rightarrow 0$ the pole approaches $2\Delta$.
Furthermore, due to the $\sqrt{4\Delta^2 - \omega^2}$ term in the numerator,
the expression is always zero at $\omega = 2\Delta$,
which leads to a negative phase change of $\pi/2$.
Thus, we expect a positive phase change of $\pi$
at the Bardasis-Schrieffer mode energy below $2\Delta$
and a negative phase change of $\pi/2$ at $2\Delta$.

As the coupling term $\chi_{\Delta B}$ contains the product of the two symmetry functions $f_\vk^s f_\vk^d$,
which are orthogonal, the term vanishes and the Higgs and Bardasis-Schrieffer modes are not coupled
but contribute individually to the response.
However, for strong pulses beyond the Gaussian level, as used in pump-probe experiments,
a finite coupling between Higgs and Bardasis-Schrieffer mode can exist \cite{Mueller2019}.
In addition, due to the symmetry function $f_\vk^d$ in $\chi_{B A^2}^{ij}$,
the coupling strength of light to the Bardasis-Schrieffer mode will depend on the polarization.
For our band structure, $\theta = \pi/4$ will correspond to the $A_{1g}$ symmetry,
such that we expend a vanishing of the $d$-wave ($B_{1g}$) Bardasis-Schrieffer mode.
This polarization sensitivity has also been discussed in \cite{Mueller2019}.

Next, we evaluate the THG response at $T=0$ numerically
for two different polarization angles $\theta = 0,\pi/4$
and show the individual contributions in Fig.~\ref{fig:higgs_bs_contributions}.
As we have anticipated, there is no coupling between Higgs and Bardasis-Schrieffer mode
and the mixed contribution is zero (not shown).
Thus, the quasiparticle (blue curve) and Higgs (red curve) response is the same as in the pure system discussed in Sec.~\ref{sec:bcs}.
The Higgs contribution is polarization independent,
while the quasiparticle contribution gets reduced for $\theta = \pi/4$.
The Bardasis-Schrieffer (green curve) mode has a strong polarization dependence.
For $\theta=0$, the expected phase behavior originating from the Bardasis-Schrieffer propagator $B(\omega)$ is visible.
At the resonance energy a positive $\pi$ phase change occurs
and at the energy of the Higgs mode a negative $\pi/2$ phase change occurs.
The intensity shows a strong peak at the Bardasis-Schrieffer resonance.
The small peak at the Higgs mode energy is a result of the susceptibility $\chi_{BA^2}$.
For $\theta=\pi/4$, the Bardasis-Schrieffer mode is not excited.

Having no coupling between the modes,
the resulting phase signature is influenced only
by the interference of the pure, individual contributions.
The total THG signal (yellow curve) consists of two resonance peaks at the original,
unrenormalized frequencies each accompanied by a positive phase change.
In between we see an antiresonance behavior with a dip in the intensity
and a negative phase change.

\begin{figure}[t]
    \centering
    \includegraphics[width=\columnwidth]{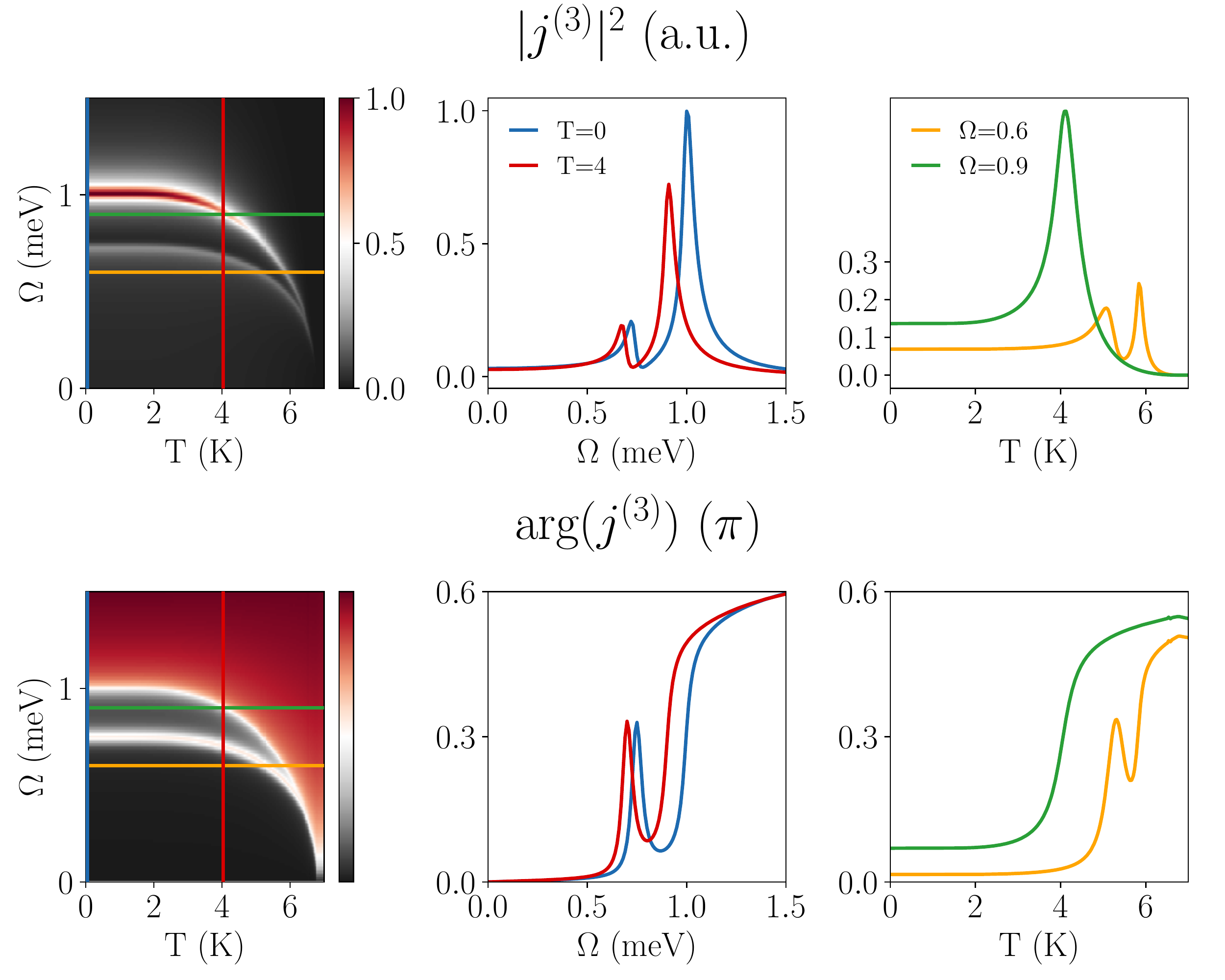}
    \caption{\label{fig:higgs_bs}%
    Intensity (top row) and phase (bottom row) of THG signal for Higgs and Bardasis-Schrieffer mode
    as function of temperature and frequency.
    The first column shows the full temperature and frequency dependence.
    The second column shows the frequency dependence for selected temperatures (vertical cuts). The third column shows the temperature dependence for selected frequencies (horizontal cuts).}
\end{figure}

In analogy to the previous section,
we also calculate the frequency and temperature dependence of the total THG signal.
The result is shown in Fig.~\ref{fig:higgs_bs}.
A before,
the temperature dependence follows roughly a quarter circle (left column).
Thus, both vertical and horizontal cuts along the frequency or temperature
axis show a similar result and the resonance and antiresonance behavior is visible in both cases.

\section{Conclusion}
\label{sec:conclusion}
In this paper,
we investigated the phase signatures of the third-harmonic (TH) signal
generated by driving superconductors with THz light.
Hereby, the phase of the oscillatory TH signal is measured
with respect to the phase of the first-harmonic.
While it is well known that resonances in the intensity of the signal
occur if the driving frequency matches with the energy of intrinsic modes,
in this paper we showed that the phase change is a robust feature as well
and additionally encodes intrinsic properties of the system.

From a classical point of view,
the resonance peak of a driven oscillator
is accompanied by a positive phase change of $\pi$.
This phase signature is more robust against damping than the peak itself,
as it is still observable even if the peak is heavily suppressed.
Coupled modes lead to antiresonance behavior
where the interference of driving force and coupling
leads to a dip in the oscillation amplitude
and a negative phase change of $\pi$.
Usually, the antiresonance is understood as a feature of a single oscillator
which is externally driven and coupled to an undriven second oscillator.
Yet, for oscillators which are both driven but uncoupled
a similar feature also occurs in the superposition of the oscillation amplitudes
as a destructive interference of two oscillations.

In a microscopic BCS model for superconductors,
the amplitude (Higgs) mode can be driven by a nonlinear, quadratic process
such that the effective driving frequency is doubled
which leads to an induced third-harmonic current.
This driving of the Higgs mode is similar to a classical oscillator, yet,
the microscopic details lead to a modification of the phase signature.

First of all, as the energy of the Higgs mode coincidences with the quasiparticle excitation energy
at the gap $2\Delta$
it is intrinsically damped.
This leads to a propagator with a square root pole structure
such that the phase change across the pole is reduced to $\pi/2$.
Secondly,
in the microscopic theory, frequency-dependent susceptibility terms occur
which govern the coupling of light to the condensate.
These terms are missing in the classical theory
and can, in principle, modify the phase signature of the observed signal.
In the clean limit, the influence of the diamagnetic coupling terms
cancel out such that the total phase response is dominated solely by the propagator.
Yet, in the dirty limit, the paramagnetic coupling terms modify the phase signature
and approximately restore the $\pi$ phase change.
Finally,
in addition to the Higgs mode response,
quasiparticles contribute to the TH signal as well
which also show a $\pi/2$ phase signature.

Considering long-ranged Coulomb interaction,
the TH response of Higgs and quasiparticles is renormalized.
Yet, one finds that these modifications do not change the phase signature.
Thus, in a BCS superconductor one can expect a $\pi/2$ phase change in the clean limit
and a $\pi$ phase change in the dirty limit at the resonance frequency.

If in addition to the Higgs mode other modes exist,
the coupled mode scenario with the antiresonance behavior is applicable.
Yet, again, microscopic details may modify the classical analysis.
To get some insight, we investigate two specific scenarios in this paper:
A coupling of the Higgs mode to a charge density wave (CDW)
and a superconducting system which additionally hosts a Bardasis-Schrieffer mode,
an amplitude fluctuation of a subleading pairing channel.

In the superconductor-CDW scenario,
the propagator of the Higgs mode itself is modified
as the energy of the Higgs mode no longer overlaps with the total energy gap
of the system which consists of the superconducting and CDW gap.
Therefore,
the propagator obtains a proper pole
and the Higgs mode becomes stable.
This restores the usual $\pi$ phase change at the Higgs mode energy.
In this system, both modes are driven by light and additionally couple to each other
such that an antiresonance effect is expected
both from the usual scenario
but also due to the interference of the individual contributions.
The structure of the analytically evaluated response exactly
corresponds to the classic coupled oscillator system.
An evaluation of the response as function of frequency and temperature
shows the antiresonance behavior and thus,
acts as fingerprint of the existence of two modes.

The second scenario with the Bardasis-Schrieffer mode serves
as an example of two uncoupled modes.
As the Bardasis-Schrieffer mode is a fluctuation of a subleading pairing channel
orthogonal to the dominant pairing channel,
the coupling element vanishes.
Nevertheless, the superposition of the individual contributions
leads to an antiresonance behavior where a dip in the intensity and a negative phase change occurs.

The considered scenarios are of course not exhaustive
but serve as a proof-of-principle how a careful examination
of the THG signal phase gives insight about the existence and nature of collective modes.
Furthermore,
these scenarios may be applicable in real systems.
The superconductor-CDW scenario is relevant for example for NbSe$_2$,
where superconductivity and CDW coexists \cite{Measson2014,Cea2014}
but might also be relevant for cuprates,
where an antiresonance behavior was observed in recent THG experiments \cite{Chu2020}.
On the other hand, Bardasis-Schrieffer modes might exists in iron-based superconductors.
As our study shows, this coupled mode scenario in a microscopic theory is
as generally applicable as in the classic theory.
Thus we expect that it can describe any other collective mode including
Leggett modes in multiband systems \cite{Murotani2017},
Josephson-Plasma modes in layered systems \cite{Gabriele2021},
or generally phonon and magnon excitations in the THz regime.

So far,
THG experiments on superconductors have only been performed
in a setup where the temperature was varied to reach the resonance condition
\cite{Matsunaga2014,Chu2020,Kovalev2020}.
As our calculation shows, the resonance and antiresonance signature of the coupled modes is also visible in this case.
Yet, to obtain a full picture, it would be necessary to obtain full temperature and frequency data to map out the temperature and frequency dependence of the modes.
A further experimental difficulty in this case is the extraction of the phase from the measured signal.
As the screening of the THz light is temperature dependent,
the first-harmonic signal might be shifted
such that a comparison of the relative TH phase could be unreliable.
While the phase change is generally more robust than resonance peaks as a signature,
strong damping may decrease or wash out the antiresonance behavior,
especially if a resonance and antiresonance are positioned close to each other.

To conclude,
we have shown that the phase of the THG signal is an interesting quantity to study
as it serves as a signature of microscopic details and coupled modes in superconductors.
It is a further step in the new field of Higgs spectroscopy
and extracting phase information in future experiments
will help to reveal more details of the investigated systems.

\section{Acknowledgements}
We are grateful to S. Kaiser, H. Chu, M. Puviani, S. Klein and J. Mitscherling for helpful discussions and comments.
We further thank the Max Planck-UBC-UTokyo Center
for Quantum Materials for fruitful collaborations and
financial support.
R.H. acknowledges the Joint-PhD program of the Univer-
sity of British Columbia and the University of Stuttgart.

\appendix

\begin{widetext}

\section{Effective action for Higgs mode}
\label{app:higgs}
We use the BCS Hamiltonian coupled to light
\begin{align}
    H(t) &= \sum_{\vk,\sigma} \epsilon_\vk c_{\vk,\sigma}^\dagger c_{\vk,\sigma}
        - \sum_{\vk,\vk'} V_{\vk,\vk'} c_{\vk,\uparrow}^\dagger
            c_{-\vk,\downarrow}^\dagger c_{-\vk',\downarrow} c_{\vk',\uparrow}
        + \frac 1 2 \sum_{\vk,\sigma} \sum_{i,j} \partial_{ij}^2\epsilon_\vk
        A_i(t) A_j(t) c_{\vk,\sigma}^\dagger c_{\vk,\sigma}
\end{align}
with electron dispersion $\epsilon_\vk = \xi_\vk - \epsilon_{\mathrm F}$
measured relative to the Fermi level
and $c_{\vk,\sigma}^\dagger$ or $c_{\vk,\sigma}$
the electron creation or annihilation operators.
We assume that the system is parity symmetric, i.e. $\epsilon_\vk = \epsilon_{-\vk}$.
The separable BCS pairing interaction is given by
$V_{\vk,\vk'} = Vf_\vk f_{\vk'}$
with pairing strength $V$ and symmetry function $f_\vk$.
The coupling to light is obtained by the expansion of the minimal coupling term
in powers of the vector potential $A$ up to second order
\begin{align}
    \epsilon_{\vk-\vec A(t)} &= \epsilon_\vk
        - \sum_i \partial_i \epsilon_\vk A_i(t)
        + \frac 1 2 \sum_{i,j} \partial_{ij}^2 \epsilon_\vk A_i(t)A_j(t)
        + \mathcal O(A(t)^3)\,.
\end{align}
Hereby, the paramagnetic term linear in $A$ vanishes due to parity symmetry,
i.e. $\partial_i \epsilon_{-\vk} = - \partial_i \epsilon_\vk$,
while only the diamagnetic term quadratic in $A$ remains
due to $\partial_{ij}^2\epsilon_{-\vk} = \partial_{ij}^2\epsilon_{\vk}$.
The partition function of the system is given by
\begin{align}
    \mathcal Z &= \int \mathcal D(c^\dagger,c) \e^{-S(c^\dagger,c)}
\end{align}
with the action in imaginary time $\tau$
\begin{align}
    S(c^\dagger, c) &= \int_0^\beta \mathrm d\tau \left(
        \sum_{\vk,\sigma} c_{\vk,\sigma}^\dagger(\tau)
        \partial_\tau c_{\vk,\sigma}(\tau) + H(\tau)
        \right)\,.
\end{align}
We decouple the pairing interaction with the help of a
Hubbard-Stratonovich transformation.
Furthermore, we allow amplitude fluctuations
via $\Delta(\tau) = \Delta + \delta\Delta(\tau)$.
Introducing the Nambu spinor $\psi_\vk^\dagger = \VV{c_{\vk,\uparrow}^\dagger}{c_{-\vk,\downarrow}}$,
the action can be written in the compact form
\begin{align}
    S(\psi^\dagger,\psi,\delta\Delta) &=
        \int_0^\beta \mathrm d\tau \left(
            \frac{|\Delta(\tau)|^2}{V}
            - \sum_{\vk} \psi_\vk^\dagger(\tau)G^{-1}(\vk,\tau) \psi_{\vk}(\tau)
        \right)
\end{align}
with the inverse Green's function
\begin{align}
    G^{-1}(\vk,\tau) &=
        -\partial_\tau \tau_0
        - \epsilon_\vk \tau_3
        + \Delta_\vk \tau_1
        - \frac 1 2\sum_{ij} \partial_{ij}^2 \epsilon_\vk A_i(t) A_j(t) \tau_3
        + \delta\Delta(\tau)f_\vk \tau_1
        \,.
\end{align}
After integration of the fermions,
one obtains in frequency representation
\begin{align}
    S(\delta\Delta,\theta,\rho) &=
        \beta \frac{\Delta^2}{V}
        + \frac 1 \beta \sum_{\im\omega_m}
            \delta\Delta(-\im\omega_m) \frac 1 V \delta\Delta(\im\omega_m)
        - \tr \ln(-G^{-1})
\end{align}
where the trace include summation over momentum and frequency and
\begin{align}
    G^{-1}(\vk,\im\omega_m,\im\omega_n) &=
        G_0^{-1}(\vk,\im\omega_m,\im\omega_n)
        - \Sigma(\vk,\im\omega_m-\im\omega_n)\,,\\
    G_0^{-1}(\vk,\im\omega_m,\im\omega_n) &= \left[
        \im\omega_m \tau_0
        - \epsilon_\vk \tau_3
        + \Delta_\vk \tau_1
        \right] \beta \delta_{\omega_m,\omega_n}\,,\\
    \Sigma(\vk,\im\omega_m-\im\omega_n) &=
        \left[
        \frac 1 2\sum_{ij} \partial_{ij}^2 \epsilon_\vk A_{ij}^2(\im\omega_m-\im\omega_n) \tau_3
        - \delta\Delta(\im\omega_m-\im\omega_n)f_\vk \tau_1
        \right]\,.
\end{align}
Expanding the logarithm for small $\Sigma$, one obtains
\begin{align}
    S(\delta\Delta) &= S_{\text{mf}}
        + S_{\text{fl}}(\delta\Delta)\,,\\
    S_{\text{mf}} &=  \beta \frac{\Delta^2}{V}
        - \tr\ln(-G_0^{-1})\,,\\
    S_{\text{fl}}(\delta\Delta) &=
        \frac 1 \beta \sum_{\im\omega_m}
            \delta\Delta(-\im\omega_m) \frac 1 V \delta\Delta(\im\omega_m)
        + \tr \sum_{n=1}^\infty
            \frac{(G_0\Sigma)^n}{n}\,.
\end{align}
Relevant for THG is the fourth order action.
Thus, we consider the second order term in the sum $\frac 1 2 \tr G_0\Sigma G_0\Sigma$
which leads to
\begin{align}
    S^{(4)}(\delta\Delta) &=
        \frac 1 2 \frac 1 \beta \sum_{\im\omega_m} \Big[
            \delta\Delta(-\im\omega_m) H^{-1}(\im\omega_m) \delta\Delta(\im\omega_m)
        - 2 \delta\Delta(-\im\omega_m) \sum_{ij} \chi_{\Delta A^2}^{ij}(\im\omega_m)A_{ij}^2(\im\omega_m)
        \notag\\&\qquad
        + \sum_{ijkl} A_{ij}^2(-\im\omega_m) A_{kl}^2(\im\omega_m)
        \chi_{A^2A^2}^{ijkl}(\im\omega_m)
        \Big]\,.
\end{align}
with the inverse Higgs propagator
\begin{align}
    H^{-1}(\im\omega_m) &= \frac 2 V + \chi_{\Delta\Delta}(\im\omega_m)
        =
        \sum_\vk f_\vk^2 \frac{4\Delta_\vk^2 - (\im\omega_n)^2}
        {E_\vk(4E_\vk^2 - (\im\omega_n)^2)} \tanh(\beta E_\vk/2)
    \label{eq:Higgs_propagator}
\end{align}
and the susceptibilities
\begin{subequations}
\begin{align}
    X_{\alpha\beta}(\vk,\vk',\im\omega_m) &=
        \frac 1 \beta \sum_{\im\omega_n} \tr[
        G_0(\vk,\im\omega_n) \tau_\alpha
        G_0(\vk',\im\omega_m+\im\omega_n) \tau_\beta
        ]\,,\\
    \chi_{\Delta\Delta}(\im\omega_m) &=
        \sum_{\vk} f_\vk^2 X_{11}(\vk,\vk,\im\omega_m)
        = - \sum_{\vk} f_\vk^2
        \frac{4\epsilon_\vk^2}
        {E_\vk(4E_\vk^2 - (\im\omega_m)^2)} \tanh(\beta E_\vk/2)\,,\\
    \chi_{\Delta A^2}^{ij}(\im\omega_m) &=
        \sum_{\vk} f_\vk
        \frac 1 2 \partial_{ij}^2 \epsilon_\vk X_{13}(\vk,\vk,\im\omega_m)
        = - \frac 1 2 \sum_{\vk} f_\vk
        \frac{\partial_{ij}^2 \epsilon_\vk 4\epsilon_\vk \Delta_\vk}
        {E_\vk(4E_\vk^2 - (\im\omega_m)^2)} \tanh(\beta E_\vk/2)\,,\\
    \chi_{A^2 A^2}^{ijkl}(\im\omega_m) &=
        \sum_{\vk}
        \frac 1 4 \partial_{ij}^2 \epsilon_\vk \partial_{kl}^2 \epsilon_\vk
        X_{33}(\vk,\vk,\im\omega_m)
        = - \frac 1 4 \sum_{\vk}
        \frac{\partial_{ij}^2 \epsilon_\vk \partial_{kl}^2 \epsilon_\vk 4\Delta_\vk^2}
        {E_\vk(4E_\vk^2 - (\im\omega_m)^2)} \tanh(\beta E_\vk/2)\,.
\end{align}
\label{eq:susceptibilities1}%
\end{subequations}
Finally, integrating the fluctuations using
\begin{align}
    \int \mathcal D(\phi^\top,\phi) \, \e^{-\frac 1 2 \frac 1\beta \sum_{\im\omega_m} \phi^\top(-\im\omega_m)M(\im\omega_m)\phi(\im\omega_m) + \phi^\top(-\im\omega_m)b(\im\omega_m) + b^\top(-\im\omega_m)\phi(\im\omega_m)} = \e^{\frac 1 2 \sum_{\im\omega_m} b^\top(-\im\omega_m) M^{-1}(\im\omega_m)b(\im\omega_m)}
\end{align}
and after analytic continuation $\im\omega_m\rightarrow \omega+\im0^+$,
one obtains
\begin{align}
    S^{(4)} = \frac 1 2 \int \mathrm d\omega
        \sum_{ijkl} \left(
        \chi_{\Delta A^2}^{ij}(-\omega)
            \chi_{\Delta A^2}^{kl}(\omega)
            H(\omega)
        + \chi_{A^2A^2}^{ijkl}(\omega)
        \right)A_{ij}^2(-\omega) A_{kl}^2(\omega)\,.
\end{align}
For the following, we will only consider linear polarized light in $x$-direction
$A(t) = A_0 \hat e_x \cos(\Omega t)$, such that we can neglect all polarization indices
and the action reads
\begin{align}
    S^{(4)} = \frac 1 2 \int \mathrm d\omega K^{(4)}(\omega) A^2(-\omega) A^2(\omega)
\end{align}
where the kernel is given by
\begin{align}
    K^{(4)} &=
        \chi_{\Delta A^2}(-\omega)
        H(\omega)
        \chi_{\Delta A^2}(\omega)
        +
        \chi_{A^2A^2}(\omega)
        \,.
\end{align}
The third-order current is computed via
\begin{align}
    j^{(3)}(3\Omega) &= - \left.\frac{\mathrm dS^{(4)}}{\mathrm dA(-\omega)}\right|_{3\Omega}
        \propto K^{(4)}(2\Omega)
\end{align}
and is proportional to the fourth-order kernel evaluated at $2\Omega$.
To analytically evaluate the momentum sums,
we assume $T=0$, $s$-wave symmetry, i.e. $f_\vk = 1$, and a constant density of states at the Fermi level
such that we can write $\sum_\vk \rightarrow \lambda \int \mathrm d\epsilon$.
We use
\begin{align}
    F(\omega) := \int \mathrm d\epsilon \frac{1}{\sqrt{\epsilon^2 + \Delta^2}(4\epsilon^2 + 4\Delta^2 - \omega^2)}
    &= \frac{2\sin^{-1}\left(\frac{\omega}{2\Delta}\right)}{\omega\sqrt{4\Delta^2 - w^2}}
    \label{eq:F_int}
\end{align}
and expand the derivative term $\sum_\vk \partial_{ij}^2\epsilon_\vk = \sum_\vk \alpha_0 + \alpha_1\epsilon_\vk$, which is valid for our band structure.
We obtain
\begin{align}
    H(\omega) &= \lambda\frac{\omega}{2\sqrt{4\Delta^2 - \omega^2}\sin^{-1}\left(\frac{\omega}{2\Delta}\right)}\,,\\
    \chi_{\Delta A^2}(\im\omega_n) &=
        - \lambda \frac \Delta 4 \alpha_1\left(
            \frac 2 V - \frac 2 \omega \sqrt{4\Delta^2 - \omega^2} \sin^{-1}\left(\frac{\omega}{2\Delta}\right)
        \right)\,,\\
    \chi_{A^2 A^2}(\im\omega_n) &=
        - \lambda \frac{\Delta^2}{2}\alpha_0^2
        \frac{2\sin^{-1}\left(\frac{\omega}{2\Delta}\right)}{\omega\sqrt{4\Delta^2 - w^2}}
        - \lambda\frac{\Delta^2}{8} \alpha_1^2
        \left(
        \frac 2 V - \frac 2 \omega \sqrt{4\Delta^2 - \omega^2} \sin^{-1}\left(\frac{\omega}{2\Delta}\right)
        \right)\,.
\end{align}

\section{Effective action with Coulomb interaction}
\label{app:coulomb}
The susceptibilities for the effective action in Eq.~\eqref{eq:eff_action_coulomb}
are given in Eq.~\eqref{eq:susceptibilities1} and
\begin{subequations}
\begin{align}
    \chi_{\Delta\rho}(\im\omega_n) &=
        \sum_{\vk} X_{13}(\vk,\vk,\im\omega_n)
        = - \Delta \sum_{\vk} \frac{4\epsilon_\vk}
        {E_\vk(4E_\vk^2 - (\im\omega_n)^2)} \tanh(\beta E_\vk/2)\,,\\
    \chi_{\rho\rho}(\im\omega_n) &=
        \sum_{\vk} X_{33}(\vk,\vk,\im\omega_n)
        = - 4\Delta^2 \sum_{\vk} \frac{1}
        {E_\vk(4E_\vk^2 - (\im\omega_n)^2)} \tanh(\beta E_\vk/2)\,,\\
    \chi_{\rho A^2}^{ij}(\im\omega_n) &=
        \sum_{\vk}
        \frac 1 2 \partial_{ij}^2 \epsilon_\vk X_{33}(\vk,\vk,\im\omega_n)
        = - \sum_{\vk}
        \frac 1 2 \partial_{ij}^2 \epsilon_\vk \frac{4\Delta^2}
        {E_\vk(4E_\vk^2 - (\im\omega_n)^2)} \tanh(\beta E_\vk/2)\,.
\end{align}
\end{subequations}

\section{Coupled oscillator}
\label{app:coupled_oscillators}
The explicit expressions for the complex amplitudes $\hat A_i$
defined in Eq.~\eqref{eq:coupled_modes_A_phi}
are given by
\begin{align}
    \hat A_1 &= \frac{F_1 P_2^{-1} - g F_2}{\tilde P_1^{-1}\tilde P_2^{-1} - g^2}
    = \frac{F_1 (\omega_2^2 - \Omega^2 + \im \gamma_2 \Omega) - gF_2}
        {
        (\omega_1^2 - \Omega^2 + \im\gamma_1\Omega)(\omega_2^2 - \Omega^2 + \im\gamma_2\Omega)
        - g^2
        }\,,\\
    \hat A_2 &= \frac{F_2 P_1^{-1} - g F_1}{\tilde P_1^{-1}\tilde P_2^{-1} - g^2}
    = \frac{F_2 (\omega_1^2 - \Omega^2 + \im \gamma_1 \Omega) - g F_1}
        {
        (\omega_1^2 - \Omega^2 + \im\gamma_1\Omega)(\omega_2^2 - \Omega^2 + \im\gamma_2\Omega)
        - g^2
        }\,.
    \label{eq:A1A2_oscillator}
\end{align}
We define
\begin{align}
    V_1 &= F_1 (\omega_2^2 - \Omega^2) - g F_2 \,, &
    V_2 &= \gamma_2 \Omega F_1 \,,\\
    V_3 &= F_2 (\omega_1^2 - \Omega^2) - g F_1 \,, &
    V_4 &= \gamma_1 \Omega F_2 \,,\\
    V_5 &= (\omega_1^2 - \Omega^2)(\omega_2^2 - \Omega^2)
        - \gamma_1 \gamma_2 \Omega^2 - g^2 \,, &
    V_6 &= \gamma_1\Omega(\omega_2^2 - \Omega^2) + \gamma_2\Omega(\omega_1^2 - \Omega^2)
\end{align}
to split the nominator and denominator in real and imaginary part
\begin{align}
    \hat A_1 &= \frac{V_1 + \im V_2}{V_5 + \im V_6}\,, &
    \hat A_2 &= \frac{V_3 + \im V_4}{V_5 + \im V_6}\,.
\end{align}
Extracting absolute value and phase and using the definition $\hat A_i = A_i\e^{-\im\phi_i}$,
finally yields for the real amplitudes $A_i$ and phases $\phi_i$
\begin{align}
    A_1(\Omega) &= \sqrt{\frac{V_1^2 + V_2^2}{V_5^2+ V_6^2}} \,, &
    \phi_1(\Omega) &= \tan^{-1}\left(\frac{V_6}{V_5}\right)
    -\tan^{-1}\left(\frac{V_2}{V_1}\right)\\
    A_2(\Omega) &= \sqrt{\frac{V_3^2 + V_4^2}{V_5^2+ V_6^2}} \,, &
    \phi_2(\Omega) &= \tan^{-1}\left(\frac{V_6}{V_5}\right)
    -\tan^{-1}\left(\frac{V_4}{V_3}\right)\,.
\end{align}
The total complex amplitude $\hat A_T$ is defined as
\begin{align}
    \hat A_T = A_T \e^{-\im\phi_T} = A_1\e^{-\im\phi_1} + A_2\e^{-\im\phi_2}\,.
\end{align}
Thus, it follows for the real amplitude $A_T$ and phase $\phi_T$
\begin{align}
    A_T &= \sqrt{A_1^2 + A_2^2 + 2A_1A_2 \cos(\phi_1-\phi_2)}\,, &
    \phi_T &= \tan^{-1}\left(
    \frac{A_1\sin\phi_1 + A_2\sin\phi_2}{A_1\cos\phi_1 + A_2\cos\phi_2}
    \right)\,.
\end{align}

\section{Higgs-CDW model}
\label{app:higgs-cdw}
The action for the BCS and phonon Hamiltonian Eq.~\eqref{eq:H_cdw} reads
\begin{align}
    S(c^\dagger,c,b^\dagger,b) &= \int_0^\beta \,\mathrm d\tau \,
    \left(
        \sum_{\vk\sigma}
            c_{\vk,\sigma}^\dagger(\tau) \partial_\tau c_{\vk,\sigma}(\tau)
            + \sum_{\vq=\pm\vQ}
                b_{\vq}^\dagger(\tau) \partial_\tau b_{\vq}(\tau)
            + H(\tau)
    \right)\,.
\end{align}
Rewriting the phonon operator as $b_\vq = \frac{1}{\sqrt{2}}(Q_\vq + \im P_{-\vq})$, integrating over the momentum variable $P_\vq$,
introducing the CDW field $D_\vk = D g_\vk$ with $D = -\sqrt{2}g Q_\vq$
and performing a Hubbard-Stratonovich transformation to decouple the superconducting pairing interaction,
one obtains
\begin{align}
    S(c^\dagger,c,\Delta^*,\Delta,D^*,D) &= \int_0^\beta \,\mathrm d\tau
    \Big(
    \sum_{\vk\sigma}
            c_{\vk,\sigma}^\dagger (\partial_\tau + \epsilon_\vk) c_{\vk,\sigma}
    - \frac{1}{g^2} D P_0^{-1}(\tau) D^*
    \notag\\&\qquad
    + \frac{\Delta^2}{V} - \sum_\vk \Delta c_{\vk\uparrow}^\dagger
        c_{-\vk\downarrow}^\dagger
        - \sum_\vk \Delta^* c_{-\vk\downarrow}c_{\vk\uparrow}
        - \sum_{\vk\sigma} D_\vk c_{\vk+\vQ,\sigma}^\dagger c_{\vk\sigma}
        - \sum_{\vk\sigma} D_\vk^* c_{\vk-\vQ,\sigma}^\dagger c_{\vk\sigma}
    \Big)
\end{align}
where the bare phonon propagator is defined as
\begin{align}
    P_0^{-1}(\tau) &= -\frac{\omega_Q^2 - \partial_\tau^2}{2\omega_Q}\,.
\end{align}
We rewrite the expression with the four-component Nambu spinor
$\psi_\vk^\dagger = (c_{\vk,\uparrow}^\dagger,c_{\vk+\vQ,\uparrow}^\dagger,
c_{-\vk,\downarrow},c_{-(\vk+\vQ),\downarrow})$
and introduce amplitude fluctuations of both fields via
$\Delta(t) = \Delta + \delta\Delta(t)$
and $D(t) = D + \delta D(t)$.
We obtain in frequency representation (see also \cite{Cea2016})
\begin{align}
    S(\psi^\dagger,\psi,\delta\Delta,\delta D) &=
        \beta\frac{\Delta^2}{V}
        + \beta \frac{D^2 w_Q}{2g^2}
        + \frac 1 \beta \sum_{\im\omega_m}
            \delta\Delta(-\im\omega_m) \frac 1 V \delta\Delta(\im\omega_m)
        - \frac{1}{g^2} \frac{1}{\beta} \sum_{\im\omega_m}
            \delta D(-\im\omega_m)
            P_0^{-1}(\im\omega_m)
            \delta D(\im\omega_m)
        \notag\\&\qquad
        - \frac{1}{\beta^2}\sum_{\im\omega_m,\im\omega_n}
        \sum_{\vk} \psi_\vk^\dagger(\im\omega_m)
            G^{-1}(\vk,\im\omega_m,\im\omega_n) \psi_{\vk}(\im\omega_n)
\end{align}
with
\begin{subequations}
\begin{align}
    G^{-1}(\vk,\im\omega_m,\im\omega_n) &=
        G_0^{-1}(\vk,\im\omega_m,\im\omega_n)
        - \Sigma(\vk,\im\omega_m-\im\omega_n)\,,\\
    G_0^{-1}(\vk,\im\omega_m,\im\omega_n) &=
        (\im\omega_m \tau_0 \otimes \sigma_0
        - \epsilon_\vk \tau_3 \otimes \sigma_3
        + \Delta \tau_1 \otimes \sigma_0 + Dg_\vk \tau_3 \otimes \sigma_1)
            \beta \delta_{\omega_m,\omega_n}\,,\\
    \Sigma(\vk,\im\omega_m-\im\omega_n) &=
    \frac 1 2\sum_{i,j} \partial_{ij}^2 \epsilon_\vk A_{ij}^2(\im\omega_m-\im\omega_n) \tau_3 \otimes \sigma_3
     - \delta\Delta(\im\omega_m-\im\omega_n) \tau_1 \otimes \sigma_0
    - \delta D(\im\omega_m-\im\omega_n)g_\vk \tau_3 \otimes \sigma_1 \,.
\end{align}
\end{subequations}
where $\tau_i$ are Pauli matrices in Nambu space and $\sigma_i$ Pauli matrices in the CDW channel.
The saddle point equations are
\begin{subequations}
\begin{align}
    \Delta &= V\sum_\vk \frac{\Delta}{E_\vk} \tanh(\beta E_\vk/2)\,,\\
    D &= \frac{4g^2}{\omega_Q} \sum_\vk g_\vk^2 \frac{D}{E_\vk} \tanh(\beta E_\vk/2)\,.
\end{align}
\end{subequations}
A diagonalization of the Hamiltonian yields the quasiparticle energy
$E_\vk = \sqrt{\epsilon_\vk^2 + \Delta^2 + |D_\vk|^2}$.
After integration of the fermions
and expansion of the logarithm as in Appendix~\ref{app:higgs},
the action is split into a mean-field part and a fluctuation part
\begin{align}
    S(\delta\Delta,\delta D) = S_{\mathrm{mf}} + S_{\mathrm{fl}}(\delta\Delta,\delta D)
\end{align}
with
\begin{subequations}
\begin{align}
    S_{\mathrm{mf}} &= \beta \frac{\Delta^2}{V}
        + \beta \frac{D^2 w_Q}{2g^2}
        - \tr \ln (-G_0^{-1})\,,\\
    S_{\mathrm{fl}}(\delta\Delta,\delta D) &=
    \frac 1 \beta \sum_{\im\omega_m}
        \delta\Delta(-\im\omega_m) \frac 1 V  \delta\Delta(\im\omega_m)
    - \frac{1}{g^2} \frac 1 \beta \sum_{\im\omega_m}
        \delta D(-\im\omega_m) P_0^{-1}(\im\omega_m) \delta D(\im\omega_m)
    + \tr \sum_{n=1}^\infty \frac{(G_0\Sigma)^n}{n}\,.
\end{align}
\end{subequations}
After evaluating the sum to second order,
one obtains the fourth order action Eq~\eqref{eq:effective_action_fluctuations}.
The Higgs propagator is defined as
\begin{align}
    H^{-1}(\im\omega_m) &= \frac 2 V + \chi_{\Delta\Delta}(\im\omega_m)
     = 2 \sum_\vk
            \frac{4\Delta^2 - (\im\omega_m)^2}
            {E_\vk(4E_\vk^2 - (\im\omega_m)^2)} \tanh(\beta E_\vk/2)
\end{align}
Analogously, the renormalized phonon propagator is defined as
\begin{align}
    P^{-1}(\im\omega_m) &= P_0^{-1}(\im\omega_m) - g^2\chi_{DD}(\im\omega_m)
     = -\frac{\Omega_Q^2 - (\im\omega_m)^2}{2\omega_Q}
\end{align}
with
\begin{align}
    \Omega_Q^2 &= \omega_Q^2 + 2\omega_Q g^2\chi_{DD}(\im\omega_m)
        = 4g^2\omega_Q \sum_{\vk} g_\vk^2
            \frac{4D_\vk^2 - (\im\omega_m)^2}
        {E_\vk(4E_\vk^2 - (\im\omega_m)^2)}
        \tanh(\beta E_\vk/2)\,.
\end{align}
Hereby, the susceptibilities are defined as
\begin{subequations}
\begin{align}
    X_{\alpha\beta\gamma\delta}(\vk,\im\omega_n) &= \frac 1 \beta \sum_{\im\omega_n}
        \tr\left[
        G_0(\vk,\im\omega_n) \tau_\alpha \otimes \sigma_\beta
        G_0(\vk,\im\omega_m+\im\omega_n) \tau_\gamma \otimes \sigma_\delta
        \right]\,,\\
    \chi_{\Delta\Delta}(\im\omega_m) &=
        \sum_{\vk} X_{1010}
        = - 8 \sum_\vk
        \frac{D_\vk^2 + \epsilon_\vk^2}
        {E_\vk(4E_\vk^2 - (\im\omega_m)^2)} \tanh(\beta E_\vk/2)\,,\\
    \chi_{\Delta D}(\im\omega_m) &=
        \sum_{\vk} g_\vk X_{1031}
        = 8 \sum_\vk g_\vk
        \frac{\Delta D_\vk}
        {E_\vk(4E_\vk^2 - (\im\omega_m)^2)} \tanh(\beta E_\vk/2)\,,\\
    \chi_{\Delta A^2}^{ij}(\im\omega_m) &=
        \sum_{\vk}
        \frac 1 2 \partial_{ij}^2\epsilon_\vk X_{1033}
        = -4 \sum_\vk \partial_{ij}^2 \epsilon_\vk
        \frac{\epsilon_\vk \Delta}
        {E_\vk(4E_\vk^2 - (\im\omega_m)^2)} \tanh(\beta E_\vk/2)\,,\\
    \chi_{DD}(\im\omega_m) &=
        \sum_{\vk} g_\vk^2 X_{3131}
        = - 8 \sum_\vk g_\vk^2
        \frac{\Delta^2 + \epsilon_\vk^2}
        {E_\vk(4E_\vk^2 - (\im\omega_m)^2)} \tanh(\beta E_\vk/2)\,,\\
    \chi_{D A^2}^{ij}(\im\omega_m) &=
        \sum_{\vk} g_\vk
        \frac 1 2 \partial_{ij}^2\epsilon_\vk X_{3133}
        = -4 \sum_\vk g_\vk \partial_{ij}^2 \epsilon_\vk
        \frac{\epsilon_\vk D_\vk}
        {E_\vk(4E_\vk^2 - (\im\omega_m)^2)} \tanh(\beta E_\vk/2)\,.
\end{align}
\end{subequations}
Integration of the amplitude fluctuations finally leads to Eq~\eqref{eq:eff_action_cdw}.

\section{Higgs-Bardasis-Schrieffer model}
\label{app:higgs-bs}
\begin{figure}[t]
    \centering
    \includegraphics[width=0.5\columnwidth]{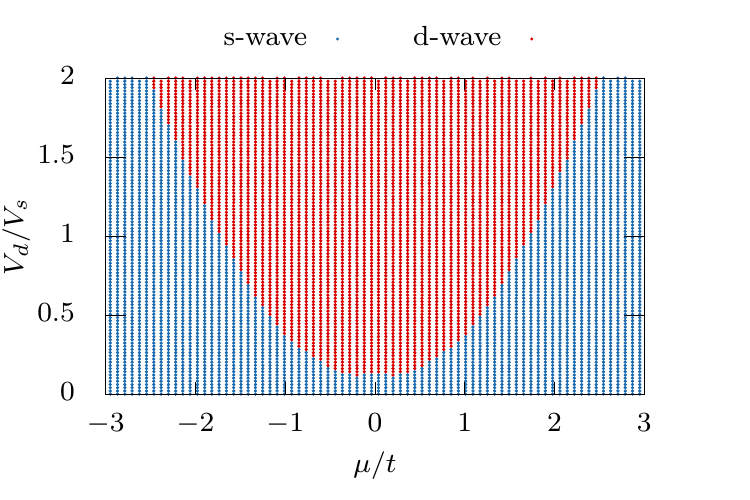}
    \caption{\label{fig:phase_diagram_bs}%
    Phase diagram showing ground state symmetry for system
    with two possible pairing channels described in Sec.~\ref{sec:higgs-bs}
    as function of chemical potential and ratio $V_d/V_s$.
    In the blue region, the $s$-wave channel is dominant
    and in the red region the $d$-wave channel.}
\end{figure}%

Using the ansatz in Eq.~\eqref{eq:bs_Vkk} for $V_{\vk\vk'}$ including the two pairing channels,
the action in imaginary time $\tau$ after decoupling of the quartic interaction
is given by
\begin{align}
    S(c^\dagger,c,\delta\Delta_l) &= \int_0^\beta \,\mathrm d\tau \, \left(
        \sum_l \frac{|\Delta_l(\tau)|^2}{V_l}
        - \sum_{\vk} \psi_\vk^\dagger(\tau) G^{-1}(\vk,\tau) \psi_{\vk}(\tau)
        \right)
\end{align}
with
\begin{align}
    G^{-1}(\vk,\tau) &= -\tau_0 \partial_\tau - h_\vk(\tau) \,,\\
    h_\vk(t) &= \left(\epsilon_\vk + \frac 1 2 \sum_{i,j} \partial_{ij}^2 \epsilon_\vk A_i(t)A_j(t)
        \right) \tau_3
        - (\Delta_s + \delta\Delta_s(t)) f_\vk^s \tau_1
        + \delta\Delta_d(t) f_\vk^d \tau_2
\end{align}
The usual Higgs mode lives in the $\tau_1$ channel,
while the Bardasis-Schrieffer mode lives in the $\tau_2$ channel.

In analogy to the previous sections,
the fermions can be integrated out which leads to
\begin{align}
    S(\delta\Delta_l) &= \beta \frac{\Delta_s^2}{V_s}
        + \frac 1 \beta \sum_{\im\omega_m} \delta\Delta_s(-\im\omega_m) \frac{1}{V_s} \delta\Delta_s(\im\omega_m)
        + \frac 1 \beta \sum_{\im\omega_m} \delta\Delta_d(-\im\omega_m) \frac{1}{V_d} \delta\Delta_d(\im\omega_m)
        -\tr \ln(-G^{-1})
\end{align}
with
\begin{align}
    G^{-1}(\vk,\im\omega_m,\im\omega_n) &=
        G_0^{-1}(\vk,\im\omega_m,\im\omega_n) - \Sigma(\vk,\im\omega_m-\im\omega_n)\,,\\
    G_0(\vk,\im\omega_m,\im\omega_n) &= (
        \im\omega_m \tau_0 - \epsilon_\vk \tau_3 + \Delta_s f_\vk^s \tau_1
    ) \beta\delta_{\omega_m,\omega_n}\,,\\
    \Sigma(\vk,\im\omega_m-\im\omega_n) &=
        \Big(
        \frac 1 2 \sum_{i,j} \partial_{ij}^2\epsilon_\vk A_{ij}^2(\im\omega_m-\im\omega_n)  \tau_3
        - \delta\Delta_s(\im\omega_m-\im\omega_n) f_\vk^s \tau_1
        + \delta\Delta_d(\im\omega_m-\im\omega_n) f_\vk^d \tau_2
        \Big)\,.
\end{align}
After expansion of the logarithm for small $\Sigma$ it follows
\begin{align}
    S(\delta\Delta_l) &= S_{\mathrm{mf}} + S_{\mathrm{fl}}(\delta\Delta_l)\,, \\
    S_{\mathrm{mf}} &= \beta \frac{\Delta_s^2}{V_s} - \tr \ln(-G_0^{-1})\,,\\
    S_{\mathrm{fl}}(\delta\Delta_l) &=
        \frac 1 \beta \sum_{\im\omega_m} \delta\Delta_s(-\im\omega_m) \frac{1}{V_s}\delta\Delta_s(\im\omega_m)
    + \frac 1 \beta \sum_{\im\omega_m} \delta\Delta_d(-\im\omega_m) \frac{1}{V_d}\delta\Delta_d(\im\omega_m)
    + \tr \sum_{n=1}^\infty \frac{(G_0\Sigma)^n}{n}\,.
\end{align}
The second-order term in the sum of the logarithm leads to the fourth-order action $S^{(4)}$
\begin{align}
    S^{(4)} &= \frac 1 2 \frac 1 \beta \sum_{\im\omega_m} \Big[
    \phi^\top(-\im\omega_m) M(\im\omega_m) \phi(\im\omega_m)
        + \phi^\top(-\im\omega_m) b(\im\omega_m)
        + b^\top(-\im\omega_m) \phi(\im\omega_m)
    \notag\\&\qquad
    + \sum_{ijkl} A_{ij}^2(-\im\omega_m)A_{kl}^2(\im\omega_m)\chi^{ijkl}_{A^2A^2}(\im\omega_m)
    \Big]
\end{align}
with $M$, $\phi$ and $b$ given in Eq.~\eqref{eq:bs_S4}.
The susceptibilities read
\begin{align}
    \chi_{\Delta\Delta}(\im\omega_m) &=
        \sum_\vk (f_\vk^s)^2 X_{11}(\vk,\im\omega_m)
        = -\sum_\vk (f_\vk^s)^2 \frac{4\epsilon_\vk^2}
            {E_\vk(4E_\vk^2 - (\im\omega_m)^2)} \tanh(\beta E_\vk/2)\\
    \chi_{BB}(\im\omega_m) &=
        \sum_\vk (f_\vk^d)^2 X_{22}(\vk,\im\omega_m)
        = -\sum_\vk (f_\vk^d)^2 \frac{4E_\vk^2}
        {E_\vk(4E_\vk^2 - (\im\omega_m)^2)} \tanh(\beta E_\vk/2)\\
    \chi_{\Delta B}(\im\omega_m) &=
        \sum_\vk f_\vk^s f_\vk^d X_{12}(\vk,\im\omega_m)
        = \sum_\vk f_\vk^s f_\vk^d \frac{2\im\epsilon_\vk (\im\omega_m)}
        {E_\vk(4E_\vk^2 - (\im\omega_m)^2)} \tanh(\beta E_\vk/2)\\
    \chi_{\Delta A^2}^{ij}(\im\omega_m) &=
        \sum_\vk \frac 1 2 f_\vk^s \partial_{ij}^2\epsilon_\vk X_{13}(\vk,\im\omega_m)
        = -\sum_\vk \frac 1 2 f_\vk^s \partial_{ij}^2\epsilon_\vk
        \frac{4\epsilon_\vk \Delta_s f_\vk^s}
        {E_\vk(4E_\vk^2 - (\im\omega_m)^2)} \tanh(\beta E_\vk/2)\\
    \chi_{B A^2}^{ij}(\im\omega_m) &=
        \sum_\vk \frac 1 2 f_\vk^d \partial_{ij}^2\epsilon_\vk X_{23}(\vk,\im\omega_m)
        = -\sum_\vk \frac 1 2 f_\vk^d \partial_{ij}^2\epsilon_\vk
        \frac{2\im\Delta_sf_\vk^s (\im\omega_m)}
        {E_\vk(4E_\vk^2 - (\im\omega_m)^2)} \tanh(\beta E_\vk/2)\\
    \chi_{A^2 A^2}^{ijkl}(\im\omega_m) &=
        \sum_\vk \frac 1 4 \partial_{ij}^2\epsilon_\vk \partial_{kl}^2\epsilon_\vk X_{33}(\vk,\im\omega_m)
        = -\sum_\vk \frac 1 4 \partial_{ij}^2\epsilon_\vk
        \partial_{kl}^2\epsilon_\vk
        \frac{4\Delta_s^2 (f_\vk^s)^2}
        {E_\vk(4E_\vk^2 - (\im\omega_m)^2)} \tanh(\beta E_\vk/2)
\end{align}
After integration of the fermions
and analytic continuation $\im\omega_m \rightarrow \omega + \im0^+$,
the action reads
\begin{align}
    S^{(4)} &= \frac 1 2 \int \mathrm d\omega
    \sum_{ijkl}
    K^{(4)}_{ijkl}(\omega) A_{ij}^2(-\omega)A_{kl}^2(\omega)
\end{align}
with the fourth-order kernel
\begin{align}
    K^{(4)}_{ijkl}(\omega) &=
    \chi_H + \chi_Q + \chi_B + \chi_M
\end{align}
where the Higgs (H), quasiparticle (Q), Bardasis-Schrieffer (B) and mixed (M) susceptibilities are given in Eq.~\eqref{eq:bs-suscept}

We consider monochromatic, linear polarized light with polarization angle $\theta$,
i.e. $\vec A(t) = A_0 \hat e \cos \Omega t$ with $\hat e^\top = \VVT{\cos\theta}{\sin\theta}$.
For the chosen tight-binding band dispersion $\epsilon_\vk = -2t(\cos k_x + \cos k_y)- \mu$, the derivative $\partial_{ij}^2\epsilon_\vk = 0$ for $i\neq j$
such that we can reduce the four polarization indices $ijkl$ to two indices $ij$, where only $\partial_{ii}^2\epsilon_\vk$ terms occur.
Thus, the action reads
\begin{align}
    S^{(4)} = \frac 1 2 \int \mathrm d\omega
    \sum_{ij} A_i^2(-\omega)A_j^2(\omega) K^{(4)}_{ij}(\omega)\,.
\end{align}
The THG current is calculated as
\begin{align}
    j^{(3)}_\alpha(3\Omega) &= - \frac{\delta S^{(4)}}{\delta A_\alpha(-\omega)}\Big|_{3\Omega}
    = - \int \mathrm d\omega'
    \sum_{j}
    A_\alpha(-\omega'+3\Omega)
    A_j^2(\omega') K^{(4)}_{\alpha j}(\omega')
    \propto \sum_j e_j^2 e_\alpha K^{(4)}_{\alpha j}(2\Omega)
\end{align}
or
\begin{align}
    \vec j^{(3)}(3\Omega) &\propto \VV{\cos^3 \theta K_{xx}^{(4)}(2\Omega) + \sin^2\theta \cos\theta K_{xy}^{(4)}(2\Omega)}
    {\cos^2\theta \sin\theta K_{yx}^{(4)}(2\Omega) + \sin^3\theta K_{yy}^{(4)}(2\Omega)}\,.
\end{align}
Thus, it follows for the current parallel to the light polarization
\begin{align}
    j_{\parallel}^{(3)}(3\Omega) &= \VV{\cos\theta}{\sin\theta} \vec j^{(3)}(3\Omega)
        = \cos^4\theta K_{xx}^{(4)}(2\Omega) + \sin^4\theta K_{yy}^{(4)}(2\Omega) + \sin^2\theta\cos^2\theta(K_{xy}^{(4)}(2\Omega)+ K_{yx}^{(4)}(2\Omega))
        \notag\\
        &= (\cos^4\theta \sin^4\theta) K_{xx}^{(4)}(2\Omega) + 2\sin^2\theta\cos^2\theta K_{xy}^{(4)}(2\Omega)
\end{align}
where we used $K_{xy}^{(4)} = K_{yx}^{(4)}$ and $K_{xx}^{(4)} = K_{yy}^{(4)}$.

We evaluate the Bardasis-Schrieffer propagator analytically for $T=0$,
and in the limit of constant density of state at the Fermi level.
We assume $f_\vk^s = 1$ and $f_\vk^d = \cos(2\varphi)$
and rewrite the sum over momentum as integral $\sum_\vk (f_\vk^d)^2 = \lambda \int \mathrm d\epsilon$
\begin{align}
    B^{-1}(\omega) &= \frac{2}{V_d} - \lambda \int \mathrm d\epsilon \frac{4\Delta^2 + 4\epsilon^2}
        {E_\vk(4E_\vk^2 - \omega^2)}\,.
\end{align}
Using Eq.~\eqref{eq:F_int} one finds
\begin{align}
    B^{-1}(\omega) &= \frac{2}{V_d} - 4\Delta^2 \lambda F(\omega) - \frac 2 V + (4\Delta^2 - \omega^2) \lambda F(\omega)
        = \frac{2}{V_d} - \frac 2 V - \omega^2 \lambda F(\omega)\,.
\end{align}

\end{widetext}


\begin{thebibliography}{36}%
\makeatletter
\providecommand \@ifxundefined [1]{%
 \@ifx{#1\undefined}
}%
\providecommand \@ifnum [1]{%
 \ifnum #1\expandafter \@firstoftwo
 \else \expandafter \@secondoftwo
 \fi
}%
\providecommand \@ifx [1]{%
 \ifx #1\expandafter \@firstoftwo
 \else \expandafter \@secondoftwo
 \fi
}%
\providecommand \natexlab [1]{#1}%
\providecommand \enquote  [1]{``#1''}%
\providecommand \bibnamefont  [1]{#1}%
\providecommand \bibfnamefont [1]{#1}%
\providecommand \citenamefont [1]{#1}%
\providecommand \href@noop [0]{\@secondoftwo}%
\providecommand \href [0]{\begingroup \@sanitize@url \@href}%
\providecommand \@href[1]{\@@startlink{#1}\@@href}%
\providecommand \@@href[1]{\endgroup#1\@@endlink}%
\providecommand \@sanitize@url [0]{\catcode `\\12\catcode `\$12\catcode
  `\&12\catcode `\#12\catcode `\^12\catcode `\_12\catcode `\%12\relax}%
\providecommand \@@startlink[1]{}%
\providecommand \@@endlink[0]{}%
\providecommand \url  [0]{\begingroup\@sanitize@url \@url }%
\providecommand \@url [1]{\endgroup\@href {#1}{\urlprefix }}%
\providecommand \urlprefix  [0]{URL }%
\providecommand \Eprint [0]{\href }%
\providecommand \doibase [0]{https://doi.org/}%
\providecommand \selectlanguage [0]{\@gobble}%
\providecommand \bibinfo  [0]{\@secondoftwo}%
\providecommand \bibfield  [0]{\@secondoftwo}%
\providecommand \translation [1]{[#1]}%
\providecommand \BibitemOpen [0]{}%
\providecommand \bibitemStop [0]{}%
\providecommand \bibitemNoStop [0]{.\EOS\space}%
\providecommand \EOS [0]{\spacefactor3000\relax}%
\providecommand \BibitemShut  [1]{\csname bibitem#1\endcsname}%
\let\auto@bib@innerbib\@empty
\bibitem [{\citenamefont {Matsunaga}\ \emph {et~al.}(2013)\citenamefont
  {Matsunaga}, \citenamefont {Hamada}, \citenamefont {Makise}, \citenamefont
  {Uzawa}, \citenamefont {Terai}, \citenamefont {Wang},\ and\ \citenamefont
  {Shimano}}]{Matsunaga2013}%
  \BibitemOpen
  \bibfield  {author} {\bibinfo {author} {\bibfnamefont {R.}~\bibnamefont
  {Matsunaga}}, \bibinfo {author} {\bibfnamefont {Y.~I.}\ \bibnamefont
  {Hamada}}, \bibinfo {author} {\bibfnamefont {K.}~\bibnamefont {Makise}},
  \bibinfo {author} {\bibfnamefont {Y.}~\bibnamefont {Uzawa}}, \bibinfo
  {author} {\bibfnamefont {H.}~\bibnamefont {Terai}}, \bibinfo {author}
  {\bibfnamefont {Z.}~\bibnamefont {Wang}},\ and\ \bibinfo {author}
  {\bibfnamefont {R.}~\bibnamefont {Shimano}},\ }\bibfield  {title} {\bibinfo
  {title} {{Higgs Amplitude Mode in the BCS Superconductors
  ${\mathrm{Nb}}_{1\mathrm{\text{\ensuremath{-}}}x}{\mathrm{Ti}}_{x}\mathbf{N}$
  Induced by Terahertz Pulse Excitation}},\ }\href
  {https://doi.org/10.1103/PhysRevLett.111.057002} {\bibfield  {journal}
  {\bibinfo  {journal} {Phys. Rev. Lett.}\ }\textbf {\bibinfo {volume} {111}},\
  \bibinfo {pages} {057002} (\bibinfo {year} {2013})}\BibitemShut {NoStop}%
\bibitem [{\citenamefont {{Matsunaga, Ryusuke and Tsuji, Naoto and Fujita,
  Hiroyuki and Sugioka, Arata and Makise, Kazumasa and Uzawa, Yoshinori and
  Terai, Hirotaka and Wang, Zhen and Aoki, Hideo and Shimano,
  Ryo}}(2014)}]{Matsunaga2014}%
  \BibitemOpen
  \bibfield  {author} {\bibinfo {author} {\bibnamefont {{Matsunaga, Ryusuke and
  Tsuji, Naoto and Fujita, Hiroyuki and Sugioka, Arata and Makise, Kazumasa and
  Uzawa, Yoshinori and Terai, Hirotaka and Wang, Zhen and Aoki, Hideo and
  Shimano, Ryo}}},\ }\bibfield  {title} {\bibinfo {title} {Light-induced
  collective pseudospin precession resonating with higgs mode in a
  superconductor},\ }\href {https://doi.org/10.1126/science.1254697} {\bibfield
   {journal} {\bibinfo  {journal} {Science}\ }\textbf {\bibinfo {volume}
  {345}},\ \bibinfo {pages} {1145–1149} (\bibinfo {year} {2014})}\BibitemShut
  {NoStop}%
\bibitem [{\citenamefont {Chu}\ \emph {et~al.}(2020)\citenamefont {Chu},
  \citenamefont {Kim}, \citenamefont {Katsumi}, \citenamefont {Kovalev},
  \citenamefont {Dawson}, \citenamefont {Schwarz}, \citenamefont {Yoshikawa},
  \citenamefont {Kim}, \citenamefont {Putzky}, \citenamefont {Li},
  \citenamefont {Raffy}, \citenamefont {Germanskiy}, \citenamefont {Deinert},
  \citenamefont {Awari}, \citenamefont {Ilyakov}, \citenamefont {Green},
  \citenamefont {Chen}, \citenamefont {Bawatna}, \citenamefont {Cristiani},
  \citenamefont {Logvenov}, \citenamefont {Gallais}, \citenamefont {Boris},
  \citenamefont {Keimer}, \citenamefont {Schnyder}, \citenamefont {Manske},
  \citenamefont {Gensch}, \citenamefont {Wang}, \citenamefont {Shimano},\ and\
  \citenamefont {Kaiser}}]{Chu2020}%
  \BibitemOpen
  \bibfield  {author} {\bibinfo {author} {\bibfnamefont {H.}~\bibnamefont
  {Chu}}, \bibinfo {author} {\bibfnamefont {M.-J.}\ \bibnamefont {Kim}},
  \bibinfo {author} {\bibfnamefont {K.}~\bibnamefont {Katsumi}}, \bibinfo
  {author} {\bibfnamefont {S.}~\bibnamefont {Kovalev}}, \bibinfo {author}
  {\bibfnamefont {R.~D.}\ \bibnamefont {Dawson}}, \bibinfo {author}
  {\bibfnamefont {L.}~\bibnamefont {Schwarz}}, \bibinfo {author} {\bibfnamefont
  {N.}~\bibnamefont {Yoshikawa}}, \bibinfo {author} {\bibfnamefont
  {G.}~\bibnamefont {Kim}}, \bibinfo {author} {\bibfnamefont {D.}~\bibnamefont
  {Putzky}}, \bibinfo {author} {\bibfnamefont {Z.~Z.}\ \bibnamefont {Li}},
  \bibinfo {author} {\bibfnamefont {H.}~\bibnamefont {Raffy}}, \bibinfo
  {author} {\bibfnamefont {S.}~\bibnamefont {Germanskiy}}, \bibinfo {author}
  {\bibfnamefont {J.-C.}\ \bibnamefont {Deinert}}, \bibinfo {author}
  {\bibfnamefont {N.}~\bibnamefont {Awari}}, \bibinfo {author} {\bibfnamefont
  {I.}~\bibnamefont {Ilyakov}}, \bibinfo {author} {\bibfnamefont
  {B.}~\bibnamefont {Green}}, \bibinfo {author} {\bibfnamefont
  {M.}~\bibnamefont {Chen}}, \bibinfo {author} {\bibfnamefont {M.}~\bibnamefont
  {Bawatna}}, \bibinfo {author} {\bibfnamefont {G.}~\bibnamefont {Cristiani}},
  \bibinfo {author} {\bibfnamefont {G.}~\bibnamefont {Logvenov}}, \bibinfo
  {author} {\bibfnamefont {Y.}~\bibnamefont {Gallais}}, \bibinfo {author}
  {\bibfnamefont {A.~V.}\ \bibnamefont {Boris}}, \bibinfo {author}
  {\bibfnamefont {B.}~\bibnamefont {Keimer}}, \bibinfo {author} {\bibfnamefont
  {A.~P.}\ \bibnamefont {Schnyder}}, \bibinfo {author} {\bibfnamefont
  {D.}~\bibnamefont {Manske}}, \bibinfo {author} {\bibfnamefont
  {M.}~\bibnamefont {Gensch}}, \bibinfo {author} {\bibfnamefont
  {Z.}~\bibnamefont {Wang}}, \bibinfo {author} {\bibfnamefont {R.}~\bibnamefont
  {Shimano}},\ and\ \bibinfo {author} {\bibfnamefont {S.}~\bibnamefont
  {Kaiser}},\ }\bibfield  {title} {\bibinfo {title} {{Phase-resolved Higgs
  response in superconducting cuprates}},\ }\href
  {https://doi.org/10.1038/s41467-020-15613-1} {\bibfield  {journal} {\bibinfo
  {journal} {Nat. Commun.}\ }\textbf {\bibinfo {volume} {11}},\ \bibinfo
  {pages} {1793} (\bibinfo {year} {2020})}\BibitemShut {NoStop}%
\bibitem [{\citenamefont {Schwarz}\ \emph {et~al.}(2020)\citenamefont
  {Schwarz}, \citenamefont {Fauseweh}, \citenamefont {Tsuji}, \citenamefont
  {Cheng}, \citenamefont {Bittner}, \citenamefont {Krull}, \citenamefont
  {Berciu}, \citenamefont {Uhrig}, \citenamefont {Schnyder}, \citenamefont
  {Kaiser},\ and\ \citenamefont {Manske}}]{Schwarz2020a}%
  \BibitemOpen
  \bibfield  {author} {\bibinfo {author} {\bibfnamefont {L.}~\bibnamefont
  {Schwarz}}, \bibinfo {author} {\bibfnamefont {B.}~\bibnamefont {Fauseweh}},
  \bibinfo {author} {\bibfnamefont {N.}~\bibnamefont {Tsuji}}, \bibinfo
  {author} {\bibfnamefont {N.}~\bibnamefont {Cheng}}, \bibinfo {author}
  {\bibfnamefont {N.}~\bibnamefont {Bittner}}, \bibinfo {author} {\bibfnamefont
  {H.}~\bibnamefont {Krull}}, \bibinfo {author} {\bibfnamefont
  {M.}~\bibnamefont {Berciu}}, \bibinfo {author} {\bibfnamefont {G.~S.}\
  \bibnamefont {Uhrig}}, \bibinfo {author} {\bibfnamefont {A.~P.}\ \bibnamefont
  {Schnyder}}, \bibinfo {author} {\bibfnamefont {S.}~\bibnamefont {Kaiser}},\
  and\ \bibinfo {author} {\bibfnamefont {D.}~\bibnamefont {Manske}},\
  }\bibfield  {title} {\bibinfo {title} {{Classification and characterization
  of nonequilibrium Higgs modes in unconventional superconductors}},\ }\href
  {https://doi.org/10.1038/s41467-019-13763-5} {\bibfield  {journal} {\bibinfo
  {journal} {Nature Communications}\ }\textbf {\bibinfo {volume} {11}},\
  \bibinfo {pages} {287} (\bibinfo {year} {2020})}\BibitemShut {NoStop}%
\bibitem [{\citenamefont {Varma}(2002)}]{Varma2002}%
  \BibitemOpen
  \bibfield  {author} {\bibinfo {author} {\bibfnamefont {C.~M.}\ \bibnamefont
  {Varma}},\ }\bibfield  {title} {\bibinfo {title} {{Higgs Boson in
  Superconductors}},\ }\href {https://doi.org/10.1023/A:1013890507658}
  {\bibfield  {journal} {\bibinfo  {journal} {J. Low Temp. Phys.}\ }\textbf
  {\bibinfo {volume} {126}},\ \bibinfo {pages} {901} (\bibinfo {year}
  {2002})}\BibitemShut {NoStop}%
\bibitem [{\citenamefont {Pekker}\ and\ \citenamefont
  {Varma}(2015)}]{Pekker2015}%
  \BibitemOpen
  \bibfield  {author} {\bibinfo {author} {\bibfnamefont {D.}~\bibnamefont
  {Pekker}}\ and\ \bibinfo {author} {\bibfnamefont {C.}~\bibnamefont {Varma}},\
  }\bibfield  {title} {\bibinfo {title} {{Amplitude/Higgs Modes in Condensed
  Matter Physics}},\ }\href
  {https://doi.org/10.1146/annurev-conmatphys-031214-014350} {\bibfield
  {journal} {\bibinfo  {journal} {Annual Review of Condensed Matter Physics}\
  }\textbf {\bibinfo {volume} {6}},\ \bibinfo {pages} {269–297} (\bibinfo
  {year} {2015})}\BibitemShut {NoStop}%
\bibitem [{\citenamefont {Papenkort}\ \emph {et~al.}(2007)\citenamefont
  {Papenkort}, \citenamefont {Axt},\ and\ \citenamefont
  {Kuhn}}]{Papenkort2007}%
  \BibitemOpen
  \bibfield  {author} {\bibinfo {author} {\bibfnamefont {T.}~\bibnamefont
  {Papenkort}}, \bibinfo {author} {\bibfnamefont {V.~M.}\ \bibnamefont {Axt}},\
  and\ \bibinfo {author} {\bibfnamefont {T.}~\bibnamefont {Kuhn}},\ }\bibfield
  {title} {\bibinfo {title} {{Coherent dynamics and pump-probe spectra of BCS
  superconductors}},\ }\href {https://doi.org/10.1103/PhysRevB.76.224522}
  {\bibfield  {journal} {\bibinfo  {journal} {Phys. Rev. B}\ }\textbf {\bibinfo
  {volume} {76}},\ \bibinfo {pages} {224522} (\bibinfo {year}
  {2007})}\BibitemShut {NoStop}%
\bibitem [{\citenamefont {Krull}\ \emph {et~al.}(2014)\citenamefont {Krull},
  \citenamefont {Manske}, \citenamefont {Uhrig},\ and\ \citenamefont
  {Schnyder}}]{Krull2014}%
  \BibitemOpen
  \bibfield  {author} {\bibinfo {author} {\bibfnamefont {H.}~\bibnamefont
  {Krull}}, \bibinfo {author} {\bibfnamefont {D.}~\bibnamefont {Manske}},
  \bibinfo {author} {\bibfnamefont {G.~S.}\ \bibnamefont {Uhrig}},\ and\
  \bibinfo {author} {\bibfnamefont {A.~P.}\ \bibnamefont {Schnyder}},\
  }\bibfield  {title} {\bibinfo {title} {{Signatures of nonadiabatic BCS state
  dynamics in pump-probe conductivity}},\ }\href
  {https://doi.org/10.1103/PhysRevB.90.014515} {\bibfield  {journal} {\bibinfo
  {journal} {Phys. Rev. B}\ }\textbf {\bibinfo {volume} {90}},\ \bibinfo
  {pages} {014515} (\bibinfo {year} {2014})}\BibitemShut {NoStop}%
\bibitem [{\citenamefont {Tsuji}\ and\ \citenamefont {Aoki}(2015)}]{Tsuji2015}%
  \BibitemOpen
  \bibfield  {author} {\bibinfo {author} {\bibfnamefont {N.}~\bibnamefont
  {Tsuji}}\ and\ \bibinfo {author} {\bibfnamefont {H.}~\bibnamefont {Aoki}},\
  }\bibfield  {title} {\bibinfo {title} {{Theory of Anderson pseudospin
  resonance with Higgs mode in superconductors}},\ }\href
  {https://doi.org/10.1103/PhysRevB.92.064508} {\bibfield  {journal} {\bibinfo
  {journal} {Phys. Rev. B}\ }\textbf {\bibinfo {volume} {92}},\ \bibinfo
  {pages} {064508} (\bibinfo {year} {2015})}\BibitemShut {NoStop}%
\bibitem [{\citenamefont {Cea}\ \emph {et~al.}(2016)\citenamefont {Cea},
  \citenamefont {Castellani},\ and\ \citenamefont {Benfatto}}]{Cea2016}%
  \BibitemOpen
  \bibfield  {author} {\bibinfo {author} {\bibfnamefont {T.}~\bibnamefont
  {Cea}}, \bibinfo {author} {\bibfnamefont {C.}~\bibnamefont {Castellani}},\
  and\ \bibinfo {author} {\bibfnamefont {L.}~\bibnamefont {Benfatto}},\
  }\bibfield  {title} {\bibinfo {title} {{Nonlinear optical effects and
  third-harmonic generation in superconductors: Cooper pairs versus Higgs mode
  contribution}},\ }\href {https://doi.org/10.1103/PhysRevB.93.180507}
  {\bibfield  {journal} {\bibinfo  {journal} {Phys. Rev. B}\ }\textbf {\bibinfo
  {volume} {93}},\ \bibinfo {pages} {180507} (\bibinfo {year}
  {2016})}\BibitemShut {NoStop}%
\bibitem [{\citenamefont {Schwarz}\ and\ \citenamefont
  {Manske}(2020)}]{Schwarz2020}%
  \BibitemOpen
  \bibfield  {author} {\bibinfo {author} {\bibfnamefont {L.}~\bibnamefont
  {Schwarz}}\ and\ \bibinfo {author} {\bibfnamefont {D.}~\bibnamefont
  {Manske}},\ }\bibfield  {title} {\bibinfo {title} {{Theory of driven Higgs
  oscillations and third-harmonic generation in unconventional
  superconductors}},\ }\href {https://doi.org/10.1103/PhysRevB.101.184519}
  {\bibfield  {journal} {\bibinfo  {journal} {Phys. Rev. B}\ }\textbf {\bibinfo
  {volume} {101}},\ \bibinfo {pages} {184519} (\bibinfo {year}
  {2020})}\BibitemShut {NoStop}%
\bibitem [{\citenamefont {Matsunaga}\ \emph {et~al.}(2017)\citenamefont
  {Matsunaga}, \citenamefont {Tsuji}, \citenamefont {Makise}, \citenamefont
  {Terai}, \citenamefont {Aoki},\ and\ \citenamefont
  {Shimano}}]{Matsunaga2017}%
  \BibitemOpen
  \bibfield  {author} {\bibinfo {author} {\bibfnamefont {R.}~\bibnamefont
  {Matsunaga}}, \bibinfo {author} {\bibfnamefont {N.}~\bibnamefont {Tsuji}},
  \bibinfo {author} {\bibfnamefont {K.}~\bibnamefont {Makise}}, \bibinfo
  {author} {\bibfnamefont {H.}~\bibnamefont {Terai}}, \bibinfo {author}
  {\bibfnamefont {H.}~\bibnamefont {Aoki}},\ and\ \bibinfo {author}
  {\bibfnamefont {R.}~\bibnamefont {Shimano}},\ }\bibfield  {title} {\bibinfo
  {title} {{Polarization-resolved terahertz third-harmonic generation in a
  single-crystal superconductor NbN: Dominance of the Higgs mode beyond the BCS
  approximation}},\ }\href {https://doi.org/10.1103/PhysRevB.96.020505}
  {\bibfield  {journal} {\bibinfo  {journal} {Phys. Rev. B}\ }\textbf {\bibinfo
  {volume} {96}},\ \bibinfo {pages} {020505} (\bibinfo {year}
  {2017})}\BibitemShut {NoStop}%
\bibitem [{\citenamefont {Murotani}\ \emph {et~al.}(2017)\citenamefont
  {Murotani}, \citenamefont {Tsuji},\ and\ \citenamefont
  {Aoki}}]{Murotani2017}%
  \BibitemOpen
  \bibfield  {author} {\bibinfo {author} {\bibfnamefont {Y.}~\bibnamefont
  {Murotani}}, \bibinfo {author} {\bibfnamefont {N.}~\bibnamefont {Tsuji}},\
  and\ \bibinfo {author} {\bibfnamefont {H.}~\bibnamefont {Aoki}},\ }\bibfield
  {title} {\bibinfo {title} {{Theory of light-induced resonances with
  collective Higgs and Leggett modes in multiband superconductors}},\ }\href
  {https://doi.org/10.1103/PhysRevB.95.104503} {\bibfield  {journal} {\bibinfo
  {journal} {Phys. Rev. B}\ }\textbf {\bibinfo {volume} {95}},\ \bibinfo
  {pages} {104503} (\bibinfo {year} {2017})}\BibitemShut {NoStop}%
\bibitem [{\citenamefont {Murotani}\ and\ \citenamefont
  {Shimano}(2019)}]{Murotani2019}%
  \BibitemOpen
  \bibfield  {author} {\bibinfo {author} {\bibfnamefont {Y.}~\bibnamefont
  {Murotani}}\ and\ \bibinfo {author} {\bibfnamefont {R.}~\bibnamefont
  {Shimano}},\ }\bibfield  {title} {\bibinfo {title} {{Nonlinear optical
  response of collective modes in multiband superconductors assisted by
  nonmagnetic impurities}},\ }\href
  {https://doi.org/10.1103/PhysRevB.99.224510} {\bibfield  {journal} {\bibinfo
  {journal} {Phys. Rev. B}\ }\textbf {\bibinfo {volume} {99}},\ \bibinfo
  {pages} {224510} (\bibinfo {year} {2019})}\BibitemShut {NoStop}%
\bibitem [{\citenamefont {Haenel}\ \emph {et~al.}(2020)\citenamefont {Haenel},
  \citenamefont {Froese}, \citenamefont {Manske},\ and\ \citenamefont
  {Schwarz}}]{Haenel2021}%
  \BibitemOpen
  \bibfield  {author} {\bibinfo {author} {\bibfnamefont {R.}~\bibnamefont
  {Haenel}}, \bibinfo {author} {\bibfnamefont {P.}~\bibnamefont {Froese}},
  \bibinfo {author} {\bibfnamefont {D.}~\bibnamefont {Manske}},\ and\ \bibinfo
  {author} {\bibfnamefont {L.}~\bibnamefont {Schwarz}},\ }\bibfield  {title}
  {\bibinfo {title} {{Time-resolved optical conductivity and Higgs oscillations
  in two-band dirty superconductors}},\ }\href
  {https://arxiv.org/abs/2012.07674} {\bibfield  {journal} {\bibinfo  {journal}
  {arXiv:2012.07674}\ } (\bibinfo {year} {2020})}\BibitemShut {NoStop}%
\bibitem [{\citenamefont {Gabriele}\ \emph {et~al.}(2021)\citenamefont
  {Gabriele}, \citenamefont {Udina},\ and\ \citenamefont
  {Benfatto}}]{Gabriele2021}%
  \BibitemOpen
  \bibfield  {author} {\bibinfo {author} {\bibfnamefont {F.}~\bibnamefont
  {Gabriele}}, \bibinfo {author} {\bibfnamefont {M.}~\bibnamefont {Udina}},\
  and\ \bibinfo {author} {\bibfnamefont {L.}~\bibnamefont {Benfatto}},\
  }\bibfield  {title} {\bibinfo {title} {{Non-linear Terahertz driving of
  plasma waves in layered cuprates}},\ }\href
  {https://doi.org/10.1038/s41467-021-21041-6} {\bibfield  {journal} {\bibinfo
  {journal} {Nature Communications}\ }\textbf {\bibinfo {volume} {12}},\
  \bibinfo {pages} {752} (\bibinfo {year} {2021})}\BibitemShut {NoStop}%
\bibitem [{\citenamefont {M\"uller}\ \emph {et~al.}(2019)\citenamefont
  {M\"uller}, \citenamefont {Volkov}, \citenamefont {Paul},\ and\ \citenamefont
  {Eremin}}]{Mueller2019}%
  \BibitemOpen
  \bibfield  {author} {\bibinfo {author} {\bibfnamefont {M.~A.}\ \bibnamefont
  {M\"uller}}, \bibinfo {author} {\bibfnamefont {P.~A.}\ \bibnamefont
  {Volkov}}, \bibinfo {author} {\bibfnamefont {I.}~\bibnamefont {Paul}},\ and\
  \bibinfo {author} {\bibfnamefont {I.~M.}\ \bibnamefont {Eremin}},\ }\bibfield
   {title} {\bibinfo {title} {{Collective modes in pumped unconventional
  superconductors with competing ground states}},\ }\href
  {https://doi.org/10.1103/PhysRevB.100.140501} {\bibfield  {journal} {\bibinfo
   {journal} {Phys. Rev. B}\ }\textbf {\bibinfo {volume} {100}},\ \bibinfo
  {pages} {140501} (\bibinfo {year} {2019})}\BibitemShut {NoStop}%
\bibitem [{\citenamefont {Cea}\ and\ \citenamefont {Benfatto}(2014)}]{Cea2014}%
  \BibitemOpen
  \bibfield  {author} {\bibinfo {author} {\bibfnamefont {T.}~\bibnamefont
  {Cea}}\ and\ \bibinfo {author} {\bibfnamefont {L.}~\bibnamefont {Benfatto}},\
  }\bibfield  {title} {\bibinfo {title} {{Nature and Raman signatures of the
  Higgs amplitude mode in the coexisting superconducting and
  charge-density-wave state}},\ }\href
  {https://doi.org/10.1103/PhysRevB.90.224515} {\bibfield  {journal} {\bibinfo
  {journal} {Phys. Rev. B}\ }\textbf {\bibinfo {volume} {90}},\ \bibinfo
  {pages} {224515} (\bibinfo {year} {2014})}\BibitemShut {NoStop}%
\bibitem [{\citenamefont {Anderson}(1963)}]{Anderson1963}%
  \BibitemOpen
  \bibfield  {author} {\bibinfo {author} {\bibfnamefont {P.~W.}\ \bibnamefont
  {Anderson}},\ }\bibfield  {title} {\bibinfo {title} {{Plasmons, Gauge
  Invariance, and Mass}},\ }\href {https://doi.org/10.1103/PhysRev.130.439}
  {\bibfield  {journal} {\bibinfo  {journal} {Phys. Rev.}\ }\textbf {\bibinfo
  {volume} {130}},\ \bibinfo {pages} {439} (\bibinfo {year}
  {1963})}\BibitemShut {NoStop}%
\bibitem [{\citenamefont {Altland}\ and\ \citenamefont
  {Simons}(2010)}]{Altland2010}%
  \BibitemOpen
  \bibfield  {author} {\bibinfo {author} {\bibfnamefont {A.}~\bibnamefont
  {Altland}}\ and\ \bibinfo {author} {\bibfnamefont {B.}~\bibnamefont
  {Simons}},\ }\href@noop {} {\emph {\bibinfo {title} {{Condensed Matter Field
  Theory}}}}\ (\bibinfo  {publisher} {Cambridge University Press},\ \bibinfo
  {year} {2010})\BibitemShut {NoStop}%
\bibitem [{\citenamefont {Jujo}(2015)}]{Jujo2015}%
  \BibitemOpen
  \bibfield  {author} {\bibinfo {author} {\bibfnamefont {T.}~\bibnamefont
  {Jujo}},\ }\bibfield  {title} {\bibinfo {title} {{Two-Photon Absorption by
  Impurity Scattering and Amplitude Mode in Conventional Superconductors}},\
  }\href {https://doi.org/10.7566/JPSJ.84.114711} {\bibfield  {journal}
  {\bibinfo  {journal} {Journal of the Physical Society of Japan}\ }\textbf
  {\bibinfo {volume} {84}},\ \bibinfo {pages} {114711} (\bibinfo {year}
  {2015})}\BibitemShut {NoStop}%
\bibitem [{\citenamefont {Silaev}(2019)}]{Silaev2019}%
  \BibitemOpen
  \bibfield  {author} {\bibinfo {author} {\bibfnamefont {M.}~\bibnamefont
  {Silaev}},\ }\bibfield  {title} {\bibinfo {title} {{Nonlinear electromagnetic
  response and Higgs-mode excitation in BCS superconductors with impurities}},\
  }\href {https://doi.org/10.1103/PhysRevB.99.224511} {\bibfield  {journal}
  {\bibinfo  {journal} {Phys. Rev. B}\ }\textbf {\bibinfo {volume} {99}},\
  \bibinfo {pages} {224511} (\bibinfo {year} {2019})}\BibitemShut {NoStop}%
\bibitem [{\citenamefont {Tsuji}\ and\ \citenamefont
  {Nomura}(2020)}]{Tsuji2020}%
  \BibitemOpen
  \bibfield  {author} {\bibinfo {author} {\bibfnamefont {N.}~\bibnamefont
  {Tsuji}}\ and\ \bibinfo {author} {\bibfnamefont {Y.}~\bibnamefont {Nomura}},\
  }\bibfield  {title} {\bibinfo {title} {{Higgs-mode resonance in third
  harmonic generation in NbN superconductors: Multiband electron-phonon
  coupling, impurity scattering, and polarization-angle dependence}},\ }\href
  {https://doi.org/10.1103/PhysRevResearch.2.043029} {\bibfield  {journal}
  {\bibinfo  {journal} {Phys. Rev. Research}\ }\textbf {\bibinfo {volume}
  {2}},\ \bibinfo {pages} {043029} (\bibinfo {year} {2020})}\BibitemShut
  {NoStop}%
\bibitem [{\citenamefont {Seibold}\ \emph {et~al.}(2021)\citenamefont
  {Seibold}, \citenamefont {Udina}, \citenamefont {Castellani},\ and\
  \citenamefont {Benfatto}}]{Seibold2021}%
  \BibitemOpen
  \bibfield  {author} {\bibinfo {author} {\bibfnamefont {G.}~\bibnamefont
  {Seibold}}, \bibinfo {author} {\bibfnamefont {M.}~\bibnamefont {Udina}},
  \bibinfo {author} {\bibfnamefont {C.}~\bibnamefont {Castellani}},\ and\
  \bibinfo {author} {\bibfnamefont {L.}~\bibnamefont {Benfatto}},\ }\bibfield
  {title} {\bibinfo {title} {{Third harmonic generation from collective modes
  in disordered superconductors}},\ }\href
  {https://doi.org/10.1103/PhysRevB.103.014512} {\bibfield  {journal} {\bibinfo
   {journal} {Phys. Rev. B}\ }\textbf {\bibinfo {volume} {103}},\ \bibinfo
  {pages} {014512} (\bibinfo {year} {2021})}\BibitemShut {NoStop}%
\bibitem [{\citenamefont {Sooryakumar}\ and\ \citenamefont
  {Klein}(1980)}]{Sooryakumar1980}%
  \BibitemOpen
  \bibfield  {author} {\bibinfo {author} {\bibfnamefont {R.}~\bibnamefont
  {Sooryakumar}}\ and\ \bibinfo {author} {\bibfnamefont {M.~V.}\ \bibnamefont
  {Klein}},\ }\bibfield  {title} {\bibinfo {title} {{Raman Scattering by
  Superconducting-Gap Excitations and Their Coupling to Charge-Density
  Waves}},\ }\href {https://doi.org/10.1103/PhysRevLett.45.660} {\bibfield
  {journal} {\bibinfo  {journal} {Phys. Rev. Lett.}\ }\textbf {\bibinfo
  {volume} {45}},\ \bibinfo {pages} {660} (\bibinfo {year} {1980})}\BibitemShut
  {NoStop}%
\bibitem [{\citenamefont {M\'easson}\ \emph {et~al.}(2014)\citenamefont
  {M\'easson}, \citenamefont {Gallais}, \citenamefont {Cazayous}, \citenamefont
  {Clair}, \citenamefont {Rodi\`ere}, \citenamefont {Cario},\ and\
  \citenamefont {Sacuto}}]{Measson2014}%
  \BibitemOpen
  \bibfield  {author} {\bibinfo {author} {\bibfnamefont {M.-A.}\ \bibnamefont
  {M\'easson}}, \bibinfo {author} {\bibfnamefont {Y.}~\bibnamefont {Gallais}},
  \bibinfo {author} {\bibfnamefont {M.}~\bibnamefont {Cazayous}}, \bibinfo
  {author} {\bibfnamefont {B.}~\bibnamefont {Clair}}, \bibinfo {author}
  {\bibfnamefont {P.}~\bibnamefont {Rodi\`ere}}, \bibinfo {author}
  {\bibfnamefont {L.}~\bibnamefont {Cario}},\ and\ \bibinfo {author}
  {\bibfnamefont {A.}~\bibnamefont {Sacuto}},\ }\bibfield  {title} {\bibinfo
  {title} {{Amplitude Higgs mode in the $2H\ensuremath{-}{\text{NbSe}}_{2}$
  superconductor}},\ }\href {https://doi.org/10.1103/PhysRevB.89.060503}
  {\bibfield  {journal} {\bibinfo  {journal} {Phys. Rev. B}\ }\textbf {\bibinfo
  {volume} {89}},\ \bibinfo {pages} {060503} (\bibinfo {year}
  {2014})}\BibitemShut {NoStop}%
\bibitem [{\citenamefont {Littlewood}\ and\ \citenamefont
  {Varma}(1981)}]{Littlewood1981}%
  \BibitemOpen
  \bibfield  {author} {\bibinfo {author} {\bibfnamefont {P.~B.}\ \bibnamefont
  {Littlewood}}\ and\ \bibinfo {author} {\bibfnamefont {C.~M.}\ \bibnamefont
  {Varma}},\ }\bibfield  {title} {\bibinfo {title} {{Gauge-Invariant Theory of
  the Dynamical Interaction of Charge Density Waves and Superconductivity}},\
  }\href {https://doi.org/10.1103/PhysRevLett.47.811} {\bibfield  {journal}
  {\bibinfo  {journal} {Phys. Rev. Lett.}\ }\textbf {\bibinfo {volume} {47}},\
  \bibinfo {pages} {811} (\bibinfo {year} {1981})}\BibitemShut {NoStop}%
\bibitem [{\citenamefont {Browne}\ and\ \citenamefont
  {Levin}(1983)}]{Browne1983}%
  \BibitemOpen
  \bibfield  {author} {\bibinfo {author} {\bibfnamefont {D.~A.}\ \bibnamefont
  {Browne}}\ and\ \bibinfo {author} {\bibfnamefont {K.}~\bibnamefont {Levin}},\
  }\bibfield  {title} {\bibinfo {title} {{Collective modes in
  charge-density-wave superconductors}},\ }\href
  {https://doi.org/10.1103/PhysRevB.28.4029} {\bibfield  {journal} {\bibinfo
  {journal} {Phys. Rev. B}\ }\textbf {\bibinfo {volume} {28}},\ \bibinfo
  {pages} {4029} (\bibinfo {year} {1983})}\BibitemShut {NoStop}%
\bibitem [{\citenamefont {Torchinsky}\ \emph {et~al.}(2013)\citenamefont
  {Torchinsky}, \citenamefont {Mahmood}, \citenamefont {Bollinger},
  \citenamefont {Božović},\ and\ \citenamefont {Gedik}}]{Torchinsky2013}%
  \BibitemOpen
  \bibfield  {author} {\bibinfo {author} {\bibfnamefont {D.~H.}\ \bibnamefont
  {Torchinsky}}, \bibinfo {author} {\bibfnamefont {F.}~\bibnamefont {Mahmood}},
  \bibinfo {author} {\bibfnamefont {A.~T.}\ \bibnamefont {Bollinger}}, \bibinfo
  {author} {\bibfnamefont {I.}~\bibnamefont {Božović}},\ and\ \bibinfo
  {author} {\bibfnamefont {N.}~\bibnamefont {Gedik}},\ }\bibfield  {title}
  {\bibinfo {title} {{Fluctuating charge-density waves in a cuprate
  superconductor}},\ }\href {https://doi.org/10.1038/nmat3571} {\bibfield
  {journal} {\bibinfo  {journal} {Nat. Mater.}\ }\textbf {\bibinfo {volume}
  {12}},\ \bibinfo {pages} {387–391} (\bibinfo {year} {2013})}\BibitemShut
  {NoStop}%
\bibitem [{\citenamefont {Hinton}\ \emph {et~al.}(2013)\citenamefont {Hinton},
  \citenamefont {Koralek}, \citenamefont {Lu}, \citenamefont {Vishwanath},
  \citenamefont {Orenstein}, \citenamefont {Bonn}, \citenamefont {Hardy},\ and\
  \citenamefont {Liang}}]{Hinton2013}%
  \BibitemOpen
  \bibfield  {author} {\bibinfo {author} {\bibfnamefont {J.~P.}\ \bibnamefont
  {Hinton}}, \bibinfo {author} {\bibfnamefont {J.~D.}\ \bibnamefont {Koralek}},
  \bibinfo {author} {\bibfnamefont {Y.~M.}\ \bibnamefont {Lu}}, \bibinfo
  {author} {\bibfnamefont {A.}~\bibnamefont {Vishwanath}}, \bibinfo {author}
  {\bibfnamefont {J.}~\bibnamefont {Orenstein}}, \bibinfo {author}
  {\bibfnamefont {D.~A.}\ \bibnamefont {Bonn}}, \bibinfo {author}
  {\bibfnamefont {W.~N.}\ \bibnamefont {Hardy}},\ and\ \bibinfo {author}
  {\bibfnamefont {R.}~\bibnamefont {Liang}},\ }\bibfield  {title} {\bibinfo
  {title} {{New collective mode in YBa${}_{2}$Cu${}_{3}$O${}_{6+x}$ observed by
  time-domain reflectometry}},\ }\href
  {https://doi.org/10.1103/PhysRevB.88.060508} {\bibfield  {journal} {\bibinfo
  {journal} {Phys. Rev. B}\ }\textbf {\bibinfo {volume} {88}},\ \bibinfo
  {pages} {060508} (\bibinfo {year} {2013})}\BibitemShut {NoStop}%
\bibitem [{\citenamefont {Bardasis}\ and\ \citenamefont
  {Schrieffer}(1961)}]{Bardasis1961}%
  \BibitemOpen
  \bibfield  {author} {\bibinfo {author} {\bibfnamefont {A.}~\bibnamefont
  {Bardasis}}\ and\ \bibinfo {author} {\bibfnamefont {J.~R.}\ \bibnamefont
  {Schrieffer}},\ }\bibfield  {title} {\bibinfo {title} {{Excitons and Plasmons
  in Superconductors}},\ }\href {https://doi.org/10.1103/PhysRev.121.1050}
  {\bibfield  {journal} {\bibinfo  {journal} {Phys. Rev.}\ }\textbf {\bibinfo
  {volume} {121}},\ \bibinfo {pages} {1050} (\bibinfo {year}
  {1961})}\BibitemShut {NoStop}%
\bibitem [{\citenamefont {Sun}\ \emph {et~al.}(2020)\citenamefont {Sun},
  \citenamefont {Fogler}, \citenamefont {Basov},\ and\ \citenamefont
  {Millis}}]{Sun2020}%
  \BibitemOpen
  \bibfield  {author} {\bibinfo {author} {\bibfnamefont {Z.}~\bibnamefont
  {Sun}}, \bibinfo {author} {\bibfnamefont {M.~M.}\ \bibnamefont {Fogler}},
  \bibinfo {author} {\bibfnamefont {D.~N.}\ \bibnamefont {Basov}},\ and\
  \bibinfo {author} {\bibfnamefont {A.~J.}\ \bibnamefont {Millis}},\ }\bibfield
   {title} {\bibinfo {title} {{Collective modes and terahertz near-field
  response of superconductors}},\ }\href
  {https://doi.org/10.1103/PhysRevResearch.2.023413} {\bibfield  {journal}
  {\bibinfo  {journal} {Phys. Rev. Research}\ }\textbf {\bibinfo {volume}
  {2}},\ \bibinfo {pages} {023413} (\bibinfo {year} {2020})}\BibitemShut
  {NoStop}%
\bibitem [{\citenamefont {Scalapino}\ and\ \citenamefont
  {Devereaux}(2009)}]{Scalapino2009}%
  \BibitemOpen
  \bibfield  {author} {\bibinfo {author} {\bibfnamefont {D.~J.}\ \bibnamefont
  {Scalapino}}\ and\ \bibinfo {author} {\bibfnamefont {T.~P.}\ \bibnamefont
  {Devereaux}},\ }\bibfield  {title} {\bibinfo {title} {{Collective $d$-wave
  exciton modes in the calculated Raman spectrum of Fe-based
  superconductors}},\ }\href {https://doi.org/10.1103/PhysRevB.80.140512}
  {\bibfield  {journal} {\bibinfo  {journal} {Phys. Rev. B}\ }\textbf {\bibinfo
  {volume} {80}},\ \bibinfo {pages} {140512} (\bibinfo {year}
  {2009})}\BibitemShut {NoStop}%
\bibitem [{\citenamefont {Maiti}\ and\ \citenamefont
  {Hirschfeld}(2015)}]{Maiti2015}%
  \BibitemOpen
  \bibfield  {author} {\bibinfo {author} {\bibfnamefont {S.}~\bibnamefont
  {Maiti}}\ and\ \bibinfo {author} {\bibfnamefont {P.~J.}\ \bibnamefont
  {Hirschfeld}},\ }\bibfield  {title} {\bibinfo {title} {{Collective modes in
  superconductors with competing $s$- and $d$-wave interactions}},\ }\href
  {https://doi.org/10.1103/PhysRevB.92.094506} {\bibfield  {journal} {\bibinfo
  {journal} {Phys. Rev. B}\ }\textbf {\bibinfo {volume} {92}},\ \bibinfo
  {pages} {094506} (\bibinfo {year} {2015})}\BibitemShut {NoStop}%
\bibitem [{\citenamefont {Maiti}\ \emph {et~al.}(2016)\citenamefont {Maiti},
  \citenamefont {Maier}, \citenamefont {B\"ohm}, \citenamefont {Hackl},\ and\
  \citenamefont {Hirschfeld}}]{Maiti2016}%
  \BibitemOpen
  \bibfield  {author} {\bibinfo {author} {\bibfnamefont {S.}~\bibnamefont
  {Maiti}}, \bibinfo {author} {\bibfnamefont {T.~A.}\ \bibnamefont {Maier}},
  \bibinfo {author} {\bibfnamefont {T.}~\bibnamefont {B\"ohm}}, \bibinfo
  {author} {\bibfnamefont {R.}~\bibnamefont {Hackl}},\ and\ \bibinfo {author}
  {\bibfnamefont {P.~J.}\ \bibnamefont {Hirschfeld}},\ }\bibfield  {title}
  {\bibinfo {title} {{Probing the Pairing Interaction and Multiple
  Bardasis-Schrieffer Modes Using Raman Spectroscopy}},\ }\href
  {https://doi.org/10.1103/PhysRevLett.117.257001} {\bibfield  {journal}
  {\bibinfo  {journal} {Phys. Rev. Lett.}\ }\textbf {\bibinfo {volume} {117}},\
  \bibinfo {pages} {257001} (\bibinfo {year} {2016})}\BibitemShut {NoStop}%
\bibitem [{\citenamefont {Kovalev}\ \emph {et~al.}(2020)\citenamefont
  {Kovalev}, \citenamefont {Dong}, \citenamefont {Shi}, \citenamefont
  {Reinhoffer}, \citenamefont {Xu}, \citenamefont {Wang}, \citenamefont {Gan},
  \citenamefont {Germanskiy}, \citenamefont {Deinert}, \citenamefont {Ilyakov},
  \citenamefont {van Loosdrecht}, \citenamefont {Wu}, \citenamefont {Wang},
  \citenamefont {Demsar},\ and\ \citenamefont {Zhe}}]{Kovalev2020}%
  \BibitemOpen
  \bibfield  {author} {\bibinfo {author} {\bibfnamefont {S.}~\bibnamefont
  {Kovalev}}, \bibinfo {author} {\bibfnamefont {T.}~\bibnamefont {Dong}},
  \bibinfo {author} {\bibfnamefont {L.-Y.}\ \bibnamefont {Shi}}, \bibinfo
  {author} {\bibfnamefont {C.}~\bibnamefont {Reinhoffer}}, \bibinfo {author}
  {\bibfnamefont {T.-Q.}\ \bibnamefont {Xu}}, \bibinfo {author} {\bibfnamefont
  {Y.}~\bibnamefont {Wang}, \bibfnamefont {Hong-Zhang abd~Wang}}, \bibinfo
  {author} {\bibfnamefont {Z.-Z.}\ \bibnamefont {Gan}}, \bibinfo {author}
  {\bibfnamefont {S.}~\bibnamefont {Germanskiy}}, \bibinfo {author}
  {\bibfnamefont {J.-C.}\ \bibnamefont {Deinert}}, \bibinfo {author}
  {\bibfnamefont {I.}~\bibnamefont {Ilyakov}}, \bibinfo {author} {\bibfnamefont
  {P.~H.~M.}\ \bibnamefont {van Loosdrecht}}, \bibinfo {author} {\bibfnamefont
  {D.}~\bibnamefont {Wu}}, \bibinfo {author} {\bibfnamefont {N.-L.}\
  \bibnamefont {Wang}}, \bibinfo {author} {\bibfnamefont {J.}~\bibnamefont
  {Demsar}},\ and\ \bibinfo {author} {\bibfnamefont {W.}~\bibnamefont {Zhe}},\
  }\bibfield  {title} {\bibinfo {title} {{Band-Selective Third-Harmonic
  Generation in Superconducting MgB2: Evidence for Higgs Amplitude Mode in the
  Dirty Limit}},\ }\href {https://arxiv.org/abs/2010.05019} {\bibfield
  {journal} {\bibinfo  {journal} {arXiv:2010.05019}\ } (\bibinfo {year}
  {2020})}\BibitemShut {NoStop}%
\end{thebibliography}
\end{document}